\documentclass[11pt,english]{article}

\usepackage[latin9]{inputenc}
\usepackage{geometry}
\usepackage{color}
\usepackage{amsmath}
\usepackage{amssymb}
\usepackage{graphicx}
\usepackage{setspace}
\makeatletter

\providecommand{\tabularnewline}{\\}

\@ifundefined{date}{}{\date{}}
\usepackage{array}
\usepackage{mathrsfs}
\usepackage{color}
\usepackage{colortbl}

\definecolor{lightgray}{gray}{0.8}

\makeatother

\usepackage{babel}
\begin{document}

\title{\textbf{A Versatility Measure for}\\
\textbf{Parametric Risk Models}}

\author{Michael R. Powers\thanks{Corresponding author; Department of Finance, School of Economics and
Management, and Schwarzman College, Tsinghua University, Beijing,
China 100084; email: powers@sem.tsinghua.edu.cn.} \ and Jiaxin Xu\thanks{Organization Department, CPC Yichang Municipal Committee, Yichang,
Hubei, China; email: jiaxinxucq@163.com.}}

\date{June 16, 2025}
\maketitle
\begin{abstract}
\begin{singlespace}
\noindent Parametric statistical methods play a central role in analyzing
risk through its underlying frequency and severity components. Given
the wide availability of numerical algorithms and high-speed computers,
researchers and practitioners often model these separate (although
possibly statistically dependent) random variables by fitting a large
number of parametric probability distributions to historical data
and then comparing goodness-of-fit statistics. However, this approach
is highly susceptible to problems of overfitting because it gives
insufficient weight to fundamental considerations of functional simplicity
and adaptability. To address this shortcoming, we propose a formal
mathematical measure for assessing the versatility of frequency and
severity distributions prior to their application. We then illustrate
this approach by computing and comparing values of the versatility
measure for a variety of probability distributions commonly used in
risk analysis.\medskip{}

\end{singlespace}
\end{abstract}

\section{Introduction}

\noindent Parametric statistical methods play a central role in analyzing
risk through its underlying frequency and severity components.\footnote{\noindent See, e.g., Klugman, Panjer, and Willmot (2019).}
Frequencies, which represent counts of the number of damage-causing
events to occur within a specified time period, generally are modeled
by nonnegative discrete random variables, $X\in\mathbb{Z}_{0}^{+}$,
whereas severities, which measure the individual damage amounts associated
with these events, typically are modeled by nonnegative continuous
random variables, $Y\in\mathbb{R}_{0}^{+}$ (which may be statistically
dependent on $X$).\footnote{\noindent The terms ``frequency'' and ``severity'' are routinely
used in the fields of operational risk management and actuarial finance,
but various alternatives (such as ``rate'', ``probability'', or
``likelihood'' for frequency and ``intensity'', ``impact'',
or ``consequence'' for severity) often appear in other scientific
disciplines (epidemiology, environmental science, etc.). } Given the broad availability of efficient numerical algorithms and
high-speed computers, researchers and practitioners often select a
parametric frequency or severity distribution for estimation purposes
by fitting a large number of potential distributions to relevant historical
data and then comparing the associated goodness-of-fit statistics.
Unfortunately, this approach can lead to model overfitting.

In the context of modeling frequency and severity data, the most obvious
type of overfitting involves selecting a distribution function with
an excessively large number of parameters, $k$, compared to the number
of observations, $n$. Such ``parametric overfitting'' is analogous
to constructing models with too many explanatory variables for a given
sample size in a conventional linear-regression setting, and can be
addressed in the same way; that is, by employing goodness-of-fit statistics
and/or tests that impose explicit penalties for each parameter estimated,
essentially balancing the costs and benefits of increasing $k$ for
a given $n$.

Another, less conspicuous type of overfitting that can occur when
modeling frequencies and severities is to select distributions whose
functional forms are characterized by \emph{idiosyncrasy} (i.e., excessive
\emph{complexity} and/or \emph{contrivance} of structure). Complexity
is analogous to including linear-regression covariates that are redundant
(i.e., linearly dependent on some subset of other explanatory variables),
whereas contrivance is analogous to including covariates that possess
no plausible relationship with the dependent variable. In linear regression,
the most straightforward means of avoiding redundant explanatory variables
is to exclude them through multicollinearity analysis. For irrelevant
explanatory variables, the solution is even simpler: they may be discarded
because of the absence of any connection to the dependent variable.\footnote{This is justified by the scientific method, to be discussed in Section
2, which generally does not support the inclusion of factors based
on empirical statistical relationships without some theoretical connection.} Unfortunately, there currently is no clear defense against idiosyncratic
distributions in risk analysis because no distribution functions defined
on $x\in\mathbb{Z}_{0}^{+}$ and $y\in\mathbb{R}_{0}^{+}$are manifestly
inappropriate, a priori, for frequencies and severities, respectively.

In the present investigation, we propose a novel approach to mitigating
this type of ``idiosyncratic overfitting'' based on the idea that,
for a fixed number of parameters, probability distributions used to
model frequencies and severities must demonstrate a reasonable degree
of functional \emph{versatility} (construed as the opposite of idiosyncrasy).
We begin, in Section 2, by developing the concept of versatility based
on underlying components of functional \emph{simplicity} and \emph{adaptability},
which in turn are motivated by the scientific method. Next, mathematical
techniques for measuring simplicity and adaptability are explored
in Sections 3 and 4, leading to a tentative measure of functional
versatility. This versatility measure is refined and formalized in
Section 5, then illustrated through application to a number of common
frequency and severity models in Section 6. Finally, in Section 7,
we offer some closing observations and conclusions.

\section{Model Versatility}

\noindent The list below provides preliminary definitions of various
terms employed in the Introduction. Although somewhat terse, these
definitions will be refined (both verbally and mathematically) in
the present and subsequent sections.
\begin{itemize}
\item Complexity - The degree of intricacy or disorder in the form of a
mathematical function.
\item Contrivance - The degree of artificiality or rigidity in the form
of a mathematical function.
\item Idiosyncrasy - The total amount of complexity and/or contrivance associated
with the form of a mathematical function.
\item Simplicity - The degree of plainness or coherence in the form of a
mathematical function (taken as the opposite of complexity).
\item Adaptability - The degree of naturalness or resiliency in the form
of a mathematical function (taken as the opposite of contrivance).
\item Versatility - The overall level of simplicity and/or adaptability
associated with the form of a mathematical function (taken as the
opposite of idiosyncrasy).
\end{itemize}
\qquad{}Clearly, the first three terms describe potential ``negative''
characteristics of parametric models, undesirable because they indicate
the presence of unnecessary complications or restrictions in the modeling
process. The last three terms, which represent the ``positive''
opposites of the first three, thus serve as appropriate objectives
of model selection. Although it might seem attractive to shorten the
list by referring to the positive concepts in terms of their negative
counterparts (e.g., replacing simplicity by ``inverse-complexity'')
or vice versa, this would create expositional difficulties. In particular,
it is natural to use idiosyncrasy to describe the model-selection
problems created by its components (complexity and contrivance), but
to connect these issues to the scientific method through the corresponding
components of versatility (simplicity and adaptability), as shown
in the following subsection.

\subsection{The Scientific Method}

\noindent In both the natural and social sciences, the scientific
method constitutes the generally accepted framework for guiding and
marking the progress of human knowledge. Despite lacking a formal
definition and possessing minor variations across research fields,
this method typically includes the following steps:\footnote{See, e.g., Powers (2012).}

(I) The statement of a null hypothesis ($H_{0}$) describing a scientific
relationship among observable quantities. This hypothesis, based on
derivations from theory\footnote{As noted in the Introduction, the scientific method generally requires
a theoretical basis for hypotheses. One motivation for this is to
avoid giving formal ``scientific'' recognition to statistically
significant results in experimental studies of paranormal phenomena
(whose causal explanations lie outside currently accepted science).
However, certain mainstream fields (e.g., medical research) routinely
overlook this requirement for pragmatic reasons, and thus satisfy
a somewhat lower standard of knowledge. (For example, experimental
studies may demonstrate the efficacy of a particular drug in treating
a certain disease without a clear understanding of the physiological
mechanism involved.)} (usually guided by inferences from previous observations), summarizes
the scientist's best understanding of the indicated relationship in
the simplest way possible.

(II) The identification of an observable phenomenon that is predicted
by the underlying hypothesis and not otherwise explicable.

(III) The design of an experiment comprising a reasonably large number
of individual observations, each of which reflects the presence/absence
of the identified phenomenon. This includes such critical issues as
identifying the statistical properties of the observations, how many
observations will be made, and the level of significance (i.e., probability
of Type 1 error, $\pi_{1}$), for rejecting the null hypothesis.

(IV) The execution and subsequent replication of the experiment described
in (III) by numerous independent researchers. This both reduces the
possibility of systematic errors associated with individual researchers
and enhances the statistical significance of results by adding more
observations.

Clearly, the above outline draws heavily from the paradigm of statistical
hypothesis testing. In fact, steps (I) through (III) essentially form
an application of hypothesis testing to a specific scientific problem,
accomplished by writing the scientific relationship under investigation
in the formal mathematical terms of a null hypothesis. For the present
study, we will view the frequency- and severity-modeling problem as
an application of the scientific method in which the null hypothesis
takes the form
\begin{equation}
H_{0}:\:W\sim F_{0}\equiv F_{W\mid\boldsymbol{a}}^{\left(0\right)}\left(w\right),
\end{equation}
where $F_{0}\equiv F_{W\mid\boldsymbol{a}}^{\left(0\right)}\left(w\right)$
represents the cumulative distribution function of either $W=X\in\mathbb{\mathbb{Z}}_{0}^{+}$
or $W=Y\in\mathbb{R}_{0}^{+}$, with unknown $k$-dimensional parameter
vector $\boldsymbol{a}=\left[a_{1},\ldots,a_{k}\right]$. The following
subsections address characteristics of $F_{0}$ that are particularly
desirable in this setting.

\subsection{Parsimony and Functional Simplicity}

\noindent As can be seen from step (I) of the outline, the scientific
method incorporates a bias in favor of forming the simplest feasible
null hypothesis consistent with theoretical derivations and/or empirical
observations. This approach reflects the principle of \emph{parsimony}
often associated with Occam\textquoteright s Razor: \textquotedblleft Entities
are not to be multiplied without necessity.\textquotedblright \footnote{Although attributed to William of Ockham (1287-1347), the first published
statement of this dictum (in Latin) was provided by the Irish Franciscan
friar and philosopher, John Punch, in 1639.} (which is sometimes rendered in modern English as ``The simplest
explanation is usually the best.''). However, this principle was
known in the West at least as early as Aristotle (384-322 BC), who
wrote (in his \emph{Posterior Analytics}): \textquotedblleft We may
assume the superiority, other things being equal, of the demonstration
which derives from fewer postulates or hypotheses.''

The principle of parsimony is embedded in many statistical model-selection
procedures through the use of goodness-of-fit statistics (such as
the Akaike and Bayesian information criteria) and/or hypothesis-testing
procedures (such as minimum Chi-squared) that include explicit penalties
for employing additional parameters. As observed in the Introduction,
such methods can address the problem of overfitting through over-parameterized
probability distributions, but not that of overfitting through idiosyncratic
(i.e., excessively complex and/or contrived) distribution functions.

To clarify these relationships, let the real-valued quantities $\mathcal{I}\left(F_{0}\right)\geq0$,
$\mathcal{C}\left(F_{0}\right)\geq0,$ and $\mathcal{T}\left(F_{0}\right)\geq0$
denote hypothetical heuristic measures of the intrinsic idiosyncrasy,
complexity, and contrivance of the distribution function $F_{0}$,
where each quantity is assumed to be independent of the number of
parameters, $k$. In addition, set $\mathcal{V}\left(F_{0}\right)=\tfrac{1}{\mathcal{I}\left(F_{0}\right)},$
$\mathcal{S}\left(F_{0}\right)=\tfrac{1}{\mathcal{C}\left(F_{0}\right)},$
and $\mathcal{A}\left(F_{0}\right)=\tfrac{1}{\mathcal{T}\left(F_{0}\right)}$
as the corresponding inverse (i.e., opposite) quantities: versatility,
simplicity, and adaptability, respectively. Then, from the discussion
of the preceding paragraph, it follows that $\mathcal{I}\left(F_{0}\right)$
may be viewed (heuristically) as a bivariate function of $\mathcal{C}\left(F_{0}\right)$
and $\mathcal{T}\left(F_{0}\right)$ (or alternatively, of $\mathcal{S}\left(F_{0}\right)$
and $\mathcal{A}\left(F_{0}\right)$); that is,
\[
\mathcal{I}\left(F_{0}\right)=\mathcal{I}\left(\mathcal{C}^{\left(+\right)},\mathcal{T}^{\left(+\right)}\right)
\]
\begin{equation}
=\mathcal{I}\left(\mathcal{S}^{\left(-\right)},\mathcal{A}^{\left(-\right)}\right),
\end{equation}
where the superscripts indicate the direction of the argument's impact
on $\mathcal{I}$ (i.e., $\left(+\right)=\textrm{positive}$ and $\left(-\right)=\textrm{negative}$).

Now let $\mathcal{P}\left(F_{0}\right)\geq0$ denote a hypothetical
real-valued measure of the parsimony of $F_{0}$, taken to be negatively
related to both its number of parameters and functional complexity
(because both quantities oppose the objective of simplicity), but
independent of functional contrivance (because unnaturally manipulated
functions can possess simple or complex structures). We then can treat
$\mathcal{P}\left(F_{0}\right)$ (heuristically) as a bivariate function
of $k$ and $\mathcal{C}\left(F_{0}\right)$ (or alternatively, of
$k$ and $\mathcal{S}\left(F_{0}\right)$):
\[
\mathcal{P}\left(F_{0}\right)=\mathcal{P}\left(k^{\left(-\right)},\mathcal{C}^{\left(-\right)}\right)
\]
\begin{equation}
=\mathcal{P}\left(k^{\left(-\right)},\mathcal{S}^{\left(+\right)}\right){\color{red},}
\end{equation}
where the superscripts again indicate the direction of the argument's
impact.

Functional simplicity, $\mathcal{S}\left(F_{0}\right)$, which both
increases parsimony and reduces idiosyncrasy, will be investigated
further in Section 3.

\subsection{Robustness and Functional Adaptability}

\noindent In addition to parsimony, the scientific method also entails
a bias in favor of the status quo, operationalized through the choice
of $\pi_{1}$ in step (III) of the above outline. By fixing the probability
of Type 1 error, $\pi_{1}=\Pr\left\{ \textrm{Reject }H_{0}\mid H_{0}\textrm{ is true}\right\} $
at a relatively low level, while not imposing any maximum value on
the probability of Type 2 error, $\pi_{2}=\Pr\left\{ \textrm{Retain }H_{0}\mid H_{0}\textrm{ is false}\right\} $,
the paradigm embodies what is sometimes called the Principle of Laplace:
``The weight of the evidence should be proportioned to the strangeness
of the facts.''\footnote{\noindent The original statement by Pierre-Simon Laplace (1749-1827)
\textendash{} \textquotedblleft {[}W{]}e ought to examine {[}inexplicable
phenomena{]} with an attention all the more scrupulous as it appears
more difficult to admit them.\textquotedblright{} (in French) \textendash{}
appeared in his ``Philosophical Essay on Probabilities'' (1814).
The ``Principle'' is a paraphrase of this admonition (in French)
first published by Swiss professor of psychology, Théodore Flournoy,
in 1899.} That is, if we currently believe the null hypothesis to be true,
then we should not abandon this belief unless confronted by compelling
evidence; and the greater the deviation in belief from the null hypothesis,
the more substantial the evidence necessary to justify it. Applying
this concept to the frequency- and severity-modeling problem, one
can see that it suggests the probability distribution under consideration
manifest \emph{robustness} in the sense of serving as a reasonable
model for a relatively large collection of possible data sets (and
thus being more resistant to rejection under hypothesis testing).

Letting $\mathcal{R}\left(F_{0}\right)\geq0$ denote a hypothetical
real-valued measure of the robustness of $F_{0}$, we will take this
quantity to be positively related to its number of parameters (more
of which permit greater accuracy of fit), negatively related to its
degree of functional contrivance (because unnaturally manipulated
functions are less adaptable to a wide range of empirical observations),
and independent of functional complexity (because unnaturally complicated
functions can enjoy various degrees of adaptability). We thus can
view $\mathcal{R}\left(F_{0}\right)$ (heuristically) as a bivariate
function of $k$ and $\mathcal{T}\left(F_{0}\right)$ (or alternatively,
of $k$ and $\mathcal{A}\left(F_{0}\right)$):
\[
\mathcal{R}\left(F_{0}\right)=\mathcal{R}\left(k^{\left(+\right)},\mathcal{T}^{\left(-\right)}\right)
\]
\begin{equation}
=\mathcal{R}\left(k^{\left(+\right)},\mathcal{A}^{\left(+\right)}\right).
\end{equation}

Functional adaptability, $\mathcal{A}\left(F_{0}\right)$, which both
increases robustness and reduces idiosyncrasy, will be studied in
Section 4.

\subsection{Functional Versatility}

\noindent Since both parsimony and robustness are negatively related
to idiosyncrasy, it might seem that seeking distribution functions,
$F_{0}$, with higher levels of these characteristics is an effective
means of avoiding functions that are idiosyncratic. However, as indicated
by equations (3) and (4), these quantities may depend explicitly on
$k$, in which case they can be increased simply by adjusting the
number of parameters, rather than by increasing functional simplicity
and adaptability (the quantities directly relevant to idiosyncrasy).

With this in mind, we will focus on functional versatility as the
inverse of idiosyncrasy:
\[
\mathcal{V}\left(F_{0}\right)=\dfrac{1}{\mathcal{I}\left(\mathcal{S}^{\left(-\right)},\mathcal{A}^{\left(-\right)}\right)}
\]
\begin{equation}
=\mathcal{V}\left(\mathcal{S}^{\left(+\right)},\mathcal{A}^{\left(+\right)}\right),
\end{equation}
and propose a formal expression for this quantity in Section 5.

The following table provides a summary of the various specialized
concepts employed in the present subsection. From this table, it is
clear that the distinctions between parsimony and simplicity, on the
one hand, and robustness and adaptability, on the other, are fairly
minor. However, it is important to differentiate between the concepts
in each pair to recognize the significant role played by the number
of distribution parameters.
\noindent \begin{center}
Table 1. Concepts Associated with Avoiding Model Overfitting
\par\end{center}

\noindent \begin{center}
\begin{tabular}{|c|c|c|}
\hline 
\textbf{Concept} & \textbf{Heuristic Measure} & \textbf{Inverse}\tabularnewline
\hline 
\hline 
Simplicity & $\mathcal{S}\left(F_{0}\right)=\dfrac{1}{\mathcal{C}\left(F_{0}\right)}$ & Complexity\tabularnewline
\hline 
Parsimony & $\mathcal{P}\left(F_{0}\right)=\mathcal{P}\left(k^{\left(-\right)},\mathcal{S}^{\left(+\right)}\right)$ & \textendash{}\tabularnewline
 & $=\mathcal{P}\left(k^{\left(-\right)},\mathcal{C}^{\left(-\right)}\right)$ & \tabularnewline
\hline 
Adaptability & $\mathcal{A}\left(F_{0}\right)=\dfrac{1}{\mathcal{T}\left(F_{0}\right)}$ & Contrivance\tabularnewline
\hline 
Robustness & $\mathcal{R}\left(F_{0}\right)=\mathcal{R}\left(k^{\left(+\right)},\mathcal{A}^{\left(+\right)}\right)$ & \textendash{}\tabularnewline
 & $=\mathcal{R}\left(k^{\left(+\right)},\mathcal{T}^{\left(-\right)}\right)$ & \tabularnewline
\hline 
Versatility & $\mathcal{V}\left(F_{0}\right)=\mathcal{V}\left(\mathcal{S}^{\left(+\right)},\mathcal{A}^{\left(+\right)}\right)$ & Idiosyncrasy\tabularnewline
 & $=\dfrac{1}{\mathcal{I}\left(F_{0}\right)}$ & \tabularnewline
\hline 
\end{tabular}
\par\end{center}

\section{Measuring Simplicity}

\subsection{Two Types of Complexity}

\noindent Recognizing that functional simplicity is the opposite of
functional complexity, it seems reasonable to measure the former by
selecting and inverting some measure of the latter, many of which
have been proposed in the research literatures of various fields:
statistical thermodynamics, information theory, algorithmic analysis,
etc. Although similarities exist among certain of these measures (especially
those based on some form of ``entropy''), differences in terminology
and perspective necessitate a precise identification of the specific
type of complexity we wish to model. In particular, we must determine
which of two distinct forms of complexity is more appropriate for
assessing probability mass functions (PMFs) and probability density
functions (PDFs) in the present context: Type A, or \emph{notational}
complexity, based on the number and arrangement of mathematical symbols
and operations required to express the function; and Type B, or \emph{morphological}
complexity, based on the function's geometric/topological properties
as a figure plotted in a two-dimensional coordinate system.

With regard to Type A, it is important to note that the complexity
of a PMF/PDF's notational form should not be conflated with notions
of artificial (and possibly tendentious) design, even if we know or
suspect a particular PMF/PDF to have been constructed by a clever
probabilist with specific objectives in mind. Certainly, functions
can be devised to achieve high levels of Kolmogorov (algorithmic)
complexity (also called Kolmogorov entropy), meaning that they cannot
be compressed into short string lengths in a mathematical coding system.\footnote{See Li and Vitányi (2019) for a general overview of Kolmogorov complexity/entropy
and Lu and Oliveira (2022) for insights into current applications.} However, such constructions necessitate intentionally injecting elements
of disorder or disorganization, since functions formed according to
an ordered plan or pattern (even if somewhat complicated) possess
less Kolmogorov complexity because they are more compressible.

In the case of Type B, we encounter an apparent paradox: PMFs/PDFs
with greater Shannon (information-theoretic) entropy (also called
Shannon information) tend to possess simpler, flatter shapes.\footnote{See Gray (2011) for a broad exposition of Shannon entropy and Nanda
and Chowdhury (2021) for a survey of major developments up to the
present. } For example, the maximum-entropy discrete distribution on $x\in\mathbb{\mathbb{Z}}_{0}^{+}$,
subject to a fixed mean, is given by the $\textrm{Geometric}\left(\varrho\right)$
PMF ($f_{X\mid\varrho}\left(x\right)=\varrho\left(1-\varrho\right)^{x}$,
for $\varrho\in\left(0,1\right)$), and the maximum-entropy continuous
distribution on $y\in\mathbb{R}_{0}^{+}$, again subject to a fixed
mean, is given by the $\textrm{Exponential}\left(\lambda\right)$
PDF ($f_{Y\mid\lambda}\left(y\right)=\lambda e^{-\lambda y}$, for
$\lambda\in\mathbb{R}^{+}$). This is attributable to the fact that,
for a given sample space and specified constraints (e.g., a fixed
mean or other moment), PMFs/PDFs providing greater Shannon entropy,
as measured by the expected ``surprise'' from a random trial, are
those that spread the distribution's probability more evenly across
the sample space. Essentially, we can say (informally, at least) that
increasing the disorder or disorganization of random draws is achieved
by decreasing the complexity of the distribution function's shape.

When choosing between the two complexity types, one can see that Type
B offers a clear advantage: it is possible to select a simpler PMF/PDF
just by favoring greater values of the distribution function's Shannon
entropy,
\begin{equation}
\textrm{H}_{F_{0}}\left(W\right)=\textrm{E}_{F_{0}}\left[-\ln\left(f_{0}\left(W\right)\right)\right],
\end{equation}
without having to analyze and evaluate the mathematical symbols and
operations involved in expressing $f_{0}$. As shown in Su et al.
(2021), numerous methods for assessing the complexity of functional
forms have been proposed in the research literature, but none is broadly
accepted, and none offers the analytical tractability of (6).

Another, more substantive, advantage of Type B complexity is demonstrated
by following example. Consider the PDF
\[
f_{Y\mid\delta,\ell}\left(y\right)=K\exp\left(-\dfrac{y}{\delta}\left[\ell+1+\sin\left(\varepsilon\eta_{1}y\right)+\sin\left(\varepsilon\eta_{2}y\right)+\cdots+\sin\left(\varepsilon\eta_{\ell}y\right)\right]\right),
\]
defined on $y\in\mathbb{R}_{0}^{+}$, where:

$\delta\in\mathbb{R}^{+}$ and $\ell\in\left\{ 1,2,\ldots\right\} $
are parameters;

$\varepsilon\in\mathbb{R}_{0}^{+}$ and the $\eta_{i}\in\left\{ 0,1\right\} $
are fixed constants; and

$K=\left[{\textstyle \int_{0}^{\infty}}\exp\left(-\tfrac{t}{\delta}\left[\ell+1+\sin\left(\varepsilon\eta_{1}t\right)+\sin\left(\varepsilon\eta_{2}t\right)+\cdots+\sin\left(\varepsilon\eta_{\ell}t\right)\right]\right)dt\right]^{-1}$
is the constant of integration.

\noindent From this formulation, it is easy to see that each of the
two conditions, (i) $\varepsilon=0$ and (ii) $\eta_{i}=0$ for all
$i$, yields the maximum-entropy Exponential PDF mentioned above.

Setting $\varepsilon=1$ and fixing the values of the parameters $\delta$
and $\ell$, one can manipulate the Type A complexity of $f_{Y\mid\delta,\ell}\left(y\right)$
by selecting the individual $\eta_{i}$ to generate more or less structured
(i.e., more or less Kolmogorov compressible) sequences, $\eta_{1},\eta_{2},\ldots,\eta_{\ell}$.\footnote{For example, if $m=10$, then the sequence $0,1,0,1,0,1,0,1,0,1$
is more structured (Kolmogorov compressible) than the sequence $0,1,1,1,0,1,0,0,1,0$. } As this occurs, Type B complexity will fluctuate independently of
Type A complexity, depending on the sequence $\eta_{1},\eta_{2},\ldots,\eta_{\ell}$
through only the sum ${\textstyle \sum_{i=1}^{\ell}\eta_{i}}$, and
achieving its lowest levels when this sum is close to 0 (so that $f_{Y\mid\delta,\ell}\left(y\right)$
approaches the Exponential PDF).

Now consider what happens as $\varepsilon\rightarrow0^{+}$. For any
fixed selection of $\eta_{1},\eta_{2},\ldots,\eta_{\ell}$, Type A
complexity remains constant as $\varepsilon$ decreases (because the
formula for $f_{Y\mid\delta,\ell}\left(y\right)$ remains essentially
the same), whereas Type B complexity decreases (because $f_{Y\mid\delta,\ell}\left(y\right)$
approaches the Exponential PDF). Clearly, this characteristic of Type
A is undesirable for our purposes because it fails to differentiate
between intrinsically simple PMFs/PDFs (such as those close to the
Exponential PDF) and substantially more complex functions. 

\subsection{Exponential Entropy and $\boldsymbol{p^{\textrm{th}}}$ Power Means}

\noindent Having concluded that Type B complexity is more appropriate
for the current context, we now turn to developing a measure of functional
simplicity beginning with the concept of Shannon entropy. Although
the intuition underlying (6) is fairly straightforward in the case
of discrete random variables, this is less true for continuous random
variables, where entropy often is motivated by analogy with the discrete
case.

For a frequency $X\sim F_{X}\left(x\right)$ defined on $x\in\mathbb{\mathbb{Z}}_{0}^{+}$,
the Shannon entropy is given by
\[
\textrm{H}_{F_{X}}\left(X\right)=\textrm{E}_{F_{X}}\left[-\ln\left(f_{X}\left(X\right)\right)\right]
\]
\begin{equation}
={\displaystyle \sum_{x=0}^{\infty}}\left(-\ln\left(f_{X}\left(x\right)\right)\right)f_{X}\left(x\right),
\end{equation}
where $-\ln\left(f_{X}\left(x\right)\right)$ denotes the surprise
associated with the random outcome $X=x$. Since $-\ln\left(\cdot\right)$
is a strictly decreasing function, this is consistent with the intuition
that outcomes with lower probability generate greater surprise. The
formal justification for the logarithmic form is that units of surprise
arising from statistically independent observations must be additive.

If the sample space of $X$ were bounded \textendash{} for example,
$x\in\left\{ 0,1,\ldots,B\right\} $ \textendash{} then it is quite
easy to show that the associated entropy,
\[
\textrm{H}_{F_{X}}\left(X\right)={\displaystyle \sum_{x=0}^{B}}\left(-\ln\left(f_{X}\left(x\right)\right)\right)f_{X}\left(x\right),
\]
would be maximized when the PMF is $\textrm{Discrete Uniform}\left(0,B\right)$;
that is, $f_{X\mid B}\left(x\right)=\tfrac{1}{B+1}$ for all $x$
and $\textrm{H}_{F_{X\mid B}}\left(X\right)=\ln\left(B+1\right)$.
However, for the sample space $x\in\mathbb{\mathbb{Z}}_{0}^{+}$,
no such result exists because the entropy diverges to $\infty$ as
$f_{X}\left(x\right)\rightarrow0^{+}$ for all $x$, but the limiting
values $f_{X}\left(x\right)=0$ do not form a proper probability distribution.
Therefore, one must introduce a further constraint, such as $\textrm{E}_{X}\left[X^{\kappa}\right]\equiv\mu_{\kappa}$,
subject to which entropy can be maximized. As noted above, if $\textrm{E}_{X}\left[X\right]\equiv\mu$,
then the maximum-entropy PMF is that of the $\textrm{Geometric}\left(\varrho\right)$
distribution, $f_{X\mid\varrho}\left(x\right)=\varrho\left(1-\varrho\right)^{x}$,
where $\varrho=\tfrac{1}{\mu+1}$ and $\textrm{H}_{F_{X\mid\varrho}}\left(X\right)=-\tfrac{1-\varrho}{\varrho}\ln\left(1-\varrho\right)-\ln\left(\varrho\right)$.

For a severity $Y\sim F_{Y}\left(y\right)$ defined on $y\in\mathbb{R}_{0}^{+}$,
the Shannon entropy becomes
\[
\textrm{H}_{F_{Y}}\left(Y\right)=\textrm{E}_{F_{Y}}\left[-\ln\left(f_{Y}\left(Y\right)\right)\right]
\]
\begin{equation}
={\displaystyle \int_{0}^{\infty}}\left(-\ln\left(f_{Y}\left(y\right)\right)\right)f_{Y}\left(y\right)dy,
\end{equation}
which often is called the differential entropy. Unfortunately, this
quantity does not share all the properties of the expression in (7).
Most notably, (8) can be negative, thereby failing to comport with
the notion of expected ``surprise'' from a random outcome. Nevertheless,
it still is possible to compare levels (higher vs. lower) of differential
entropy among probability distributions, and to speak of maximum-entropy
distributions.

If the sample space of $Y$ were bounded (e.g., $y\in\left[0,B\right)$),
then, in a manner analogous to the discrete case, the associated entropy
would be maximized for the $\textrm{Uniform}\left(0,B\right)$ PDF
with $f_{Y\mid B}\left(y\right)=\tfrac{1}{B}$ for all $y$ and $\textrm{H}_{F_{Y\mid B}}\left(Y\right)=\ln\left(B\right)$.
For $y\in\mathbb{R}_{0}^{+}$, one again must introduce a further
constraint (e.g., $\textrm{E}_{Y}\left[Y^{\kappa}\right]\equiv\mu_{\kappa}$),
as in the discrete case. As mentioned before, if $\textrm{E}_{Y}\left[Y\right]\equiv\mu$,
then the maximum-entropy PDF is that of the $\textrm{Exponential}\left(\lambda\right)$
distribution, $f_{Y\mid\lambda}\left(y\right)=\lambda e^{-\lambda y}$,
where $\lambda=\tfrac{1}{\mu}$ and $\textrm{H}_{F_{Y\mid\lambda}}\left(Y\right)=1-\ln\left(\lambda\right)$.
For both the $\textrm{Uniform}\left(0,B\right)$ and $\textrm{Exponential}\left(\lambda\right)$
distributions, the differential entropy is negative for certain parameter
values: $B\in\left(0,1\right)$ in the former case, and $\lambda\in\left(e,\infty\right)$
in the latter.

Given that we seek a nonnegative measure of functional simplicity,
$\mathcal{S}\left(F_{0}\right)$, the Shannon entropy cannot be used
directly. However, exponentiating this quantity yields a nonnegative
alternative: the exponentiated entropy,
\begin{equation}
e^{\textrm{H}_{F_{0}}\left(W\right)}=\exp\left(\textrm{E}_{F_{0}}\left[-\ln\left(f_{0}\left(W\right)\right)\right]\right),
\end{equation}
which preserves the order of $\textrm{H}_{F_{0}}\left(W\right)$ across
different distribution functions. Moreover, this new quantity, like
the Shannon entropy itself, does not depend directly on the number
of parameters ($k$) characterizing $F_{0}$.

Rewriting (9) as
\[
\exp\left(\textrm{E}_{F_{0}}\left[\ln\left(\dfrac{1}{f_{0}\left(W\right)}\right)\right]\right)=\underset{p\rightarrow0^{+}}{\lim}\left(\textrm{E}_{F_{0}}\left[\left|\dfrac{1}{f_{0}\left(W\right)}\right|^{p}\right]\right)^{1/p}
\]
reveals that the exponentiated entropy is essentially the geometric
mean (or $0^{\textrm{th}}$ power mean) of $\tfrac{1}{f_{0}\left(W\right)}$
taken over $F_{0}\left(w\right)$. This is noteworthy because a closely
related quantity, $\left(\textrm{E}_{F_{0}}\left[f_{0}\left(W\right)\right]\right)^{-1}$,
the inverse of the Herfindahl-Hirschman index\footnote{The Herfindahl-Hirschman index often is used as a measure of market
concentration in regulatory economics. Although usually defined for
discrete probability distributions, its extension to the continuous
case is straightforward.} (or $1^{\textrm{st}}$ power mean of $f_{0}\left(W\right)$), constitutes
a natural measure of the dispersion of the probability distribution
(see, e.g., Powers and Powers, 2015), for which greater values indicate
a simpler, flatter PMF/PDF.

Recognizing that the $p^{\textrm{th}}$ power mean of $\tfrac{1}{f_{0}\left(W\right)}$
reasonably quantifies the distribution's simplicity for all $p\in\mathbb{R}_{0}^{+}$,
we tentatively consider all
\begin{equation}
\mathcal{S}\left(F_{0}\right)=\left(\textrm{E}_{F_{0}}\left[\left|\dfrac{1}{f_{0}\left(W\right)}\right|^{p}\right]\right)^{1/p}
\end{equation}
as potential measures of functional simplicity.

\section{Measuring Adaptability}

\subsection{Fisher Information}

\noindent For the moment, assume $F_{0}\equiv F_{W\mid a}^{\left(0\right)}\left(w\right)$,
where $a$ is a scalar parameter belonging to some open set $\mathbb{A}\subseteq\mathbb{R}$.
To evaluate the functional adaptability of this probability distribution,
it is natural to consider the impact of $a$ on the corresponding
PMF or PDF, $f_{0}\left(w\right)$ \textendash{} an effect that is
measured by the Fisher information associated with a single observation
of $W$:
\[
\textrm{I}_{F_{0}}\left(a\right)=\textrm{E}_{F_{0}}\left[\left(\dfrac{\partial\ln\left(f_{0}\left(W\right)\right)}{\partial a}\right)^{2}\right]
\]
\begin{equation}
=-\textrm{E}_{F_{0}}\left[\dfrac{\partial^{2}\ln\left(f_{0}\left(W\right)\right)}{\partial a^{2}}\right],
\end{equation}
At a given value of $a$, $\textrm{I}_{F_{0}}\left(a\right)\in\mathbb{R}^{+}$
quantifies the extent by which changes in the parameter modify the
shape of the PMF or PDF, on average.\footnote{See Lehmann and Casella (1998) for a detailed presentation of Fisher
information. The expressions in (11) are well defined if, for all
$a\in\mathbb{A}$ apart from subsets of measure 0 (i.e., ``almost
everywhere''): (1) the sample space of $W\sim F_{0}\left(w\right)$
(i.e., the set $\mathbb{W}$ such that $w\in\mathbb{W}\Longrightarrow f_{0}\left(w\right)>0$)
is invariant over $a$; (2) $\tfrac{\partial f_{0}\left(w\right)}{\partial a}$
and $\tfrac{\partial^{2}f_{0}\left(w\right)}{\partial a^{2}}$ exist
and are integrable with respect to $F_{0}\left(w\right)$; and (3)
${\textstyle \int}_{\mathbb{W}}f_{0}\left(w\right)dF_{0}\left(w\right)$
is twice-differentiable under the integral sign with respect to $a$.}

Like Shannon entropy, Fisher information is additive for independent
random variables.\footnote{Although Fisher information and Shannon entropy (information) possess
certain similarities, they are designed to measure different types
of ``information'' in different contexts, and therefore are not
directly comparable. Nevertheless, they do share a mathematical connection
through the calculation of relative entropy (Kullback-Leibler divergence).
See, e.g., Cover and Thomas (2006). } In particular, for a sample of $n$ independent and identically distributed
(IID) observations, $\boldsymbol{W}=\left[W_{1},W_{2},\ldots,W_{n}\right]$,
(11) aggregates to
\[
\textrm{I}_{F_{0}}\left(a,n\right)={\displaystyle \sum_{i=1}^{n}}\textrm{E}_{F_{0}}\left[\left(\dfrac{\partial\ln\left(f_{0}\left(W_{i}\right)\right)}{\partial a}\right)^{2}\right]
\]
\begin{equation}
=n\textrm{E}_{F_{0}}\left[\left(\dfrac{\partial\ln\left(f_{0}\left(W_{i}\right)\right)}{\partial a}\right)^{2}\right]
\end{equation}
\begin{equation}
=-n\textrm{E}_{F_{0}}\left[\dfrac{\partial^{2}\ln\left(f_{0}\left(W_{i}\right)\right)}{\partial a^{2}}\right],
\end{equation}
and can be viewed as the amount of knowledge one can learn about the
parameter $a$ by observing a random sample of size $n$. Although
mathematically equivalent, the expressions in (12) and (13) offer
different perspectives on this quantity. 

In (12), the square of the first derivative captures the degree to
which marginal changes in the parameter affect the shape of the PMF/PDF
(at a fixed value of $a$), computed as an average over all possible
values of $W$. In this case, a greater impact on the PMF/PDF, and
thus the underlying probability distribution, implies that observations
of $W$ tend to provide greater information about how the parameter
characterizes the distribution. Since the PMF/PDF is transformed to
the logarithmic scale, information grows in direct proportion to the
sample size.

The expression in (13) reflects how steep, or pointed, the PMF/PDF
is with respect to marginal changes in the parameter (at a fixed value
of $a$) computed as an average over all possible values of $W$.
Here, greater steepness (which is consistent with a greater impact
of the parameter on the PMF/PDF) shows that the benefit of greater
information about the parameter is accompanied by a notable shortcoming:
small errors in estimating the true value of the parameter can lead
to substantial errors in inference about the underlying probability
distribution.

The interplay of negative and positive implications of the Fisher
information becomes apparent in the context of the Cramér-Rao lower
bound on the asymptotic variance of any consistent estimator, $\hat{a}_{n}$,
of $a$:\footnote{A consistent estimator, $\hat{a}_{n}$, converges in probability to
the true parameter, $a$. This is denoted by $\hat{a}_{n}\overset{\mathcal{P}}{\rightarrow}a$;
that is, for any $\epsilon>0$, $\underset{n\rightarrow\infty}{\lim}\Pr\left\{ \left|\hat{a}_{n}-a\right|>\epsilon\right\} =0$.}
\[
\textrm{Var}{}_{F_{0}}\left[\hat{a}_{n}\right]\geq\dfrac{1}{\textrm{I}_{F_{0}}\left(a,n\right)}.
\]
Since increasing the sample size, $n$, results in a proportional
increase in the Fisher information, it follows that the variance of
any consistent estimator of $a$ will tend to decrease over $n$,
while bounded below by the inverse of the Fisher information. Thus,
in terms of changes in the sample size, greater information is associated
with a smaller variance, which makes intuitive sense. 

However, for a fixed sample size, one also can think of the relationship
between the variance of the estimator and the Fisher information as
permitting a tradeoff: for different parameterizations of the PMF/PDF,
a parameterization that reduces the variance will imply an increase
in the Fisher information. In this case, the increase in information
is not beneficial because it means that any error in the estimate
of $a$ that does occur will magnify certain errors in estimation
of the relevant PMF/PDF.

To illustrate, consider an $\textrm{Exponential}\left(\lambda\right)$
random variable $Y$ with parameter $\lambda=6$, for which the asymptotic
variance of the sample parameter, $\hat{\lambda}_{n}=\tfrac{1}{\bar{Y}_{n}}$,
is $\tfrac{\lambda^{2}}{n}=\tfrac{36}{n}$. If we change the parameterization
of $Y$ by writing the PDF in terms of the mean, $\theta=\tfrac{1}{\lambda}=\tfrac{1}{6}$,
then the asymptotic variance of the sample estimator, $\hat{\theta}_{n}=\bar{Y}_{n}$,
becomes much smaller: $\tfrac{\theta^{2}}{n}=\tfrac{1}{36n}$. However,
using the latter estimator also causes many common calculations based
on the estimate of the PDF, such as
\[
\Pr\left\{ Y>1\right\} =e^{-\lambda}=e^{-1/\theta},
\]
to become much more sensitive to errors in the parameter estimator.
For example, assuming an error of $+0.1000$ in $\hat{\lambda}_{n}$,
we would estimate $\Pr\left\{ Y>1\right\} $ as $0.0022$, which falls
short of the true value\linebreak{}
 ($\Pr\left\{ Y>1\right\} \approx0.0025$) by $0.0003$. To achieve
a comparably small deviation in the estimate of $\Pr\left\{ Y>1\right\} $
under the alternative parameterization would require the absolute
error in $\hat{\theta}_{n}$ be no greater than
\[
\max\left\{ \left|\dfrac{1}{-\ln\left(0.0025-0.0003\right)}-\dfrac{1}{6}\right|,\left|\dfrac{1}{-\ln\left(0.0025+0.0003\right)}-\dfrac{1}{6}\right|\right\} \approx\max\left\{ \left|-0.0032\right|,\left|0.0035\right|\right\} 
\]
\[
=0.0035.
\]

Computation of the Fisher information frequently arises in the context
of maximum-likelihood estimation (MLE) because the estimator generated,
$\hat{a}_{n}$, is (under the regularity conditions identified in
Footnote 12) asymptotically Normal,\footnote{An asymptotically Normal estimator, $\hat{a}_{n}$, converges in distribution
to the Normal distribution with mean $a$ (the true parameter) and
variance equal to the Cramér-Rao lower bound. This is denoted by $\hat{a}_{n}\overset{\mathcal{D}}{\rightarrow}\textrm{Normal}\left(a,\:\tfrac{1}{\textrm{I}_{F_{0}}\left(a,n\right)}\right)$;
that is, $\underset{n\rightarrow\infty}{\lim}\left|\Pr\left\{ \tfrac{\hat{a}_{n}-a}{\sqrt{1/\textrm{I}_{F_{0}}\left(a,n\right)}}\leq t\right\} -\Phi\left(t\right)\right|=0$
for all $t\in\mathbb{R}$, where $\Phi\left(\cdot\right)$ is the
CDF of the Standard Normal distribution.} permitting its variance to be estimated well by the inverse of the
Fisher information; that is,
\[
\textrm{Var}{}_{F_{0}}\left[\hat{a}_{n}\right]\approx\dfrac{1}{\textrm{I}_{F_{0}}\left(\hat{a}_{n},n\right)}
\]
for large $n,$ where $\textrm{I}_{F_{0}}\left(\hat{a}_{n},n\right)$
denotes an estimator of $\textrm{I}_{F_{0}}\left(a,n\right)$ obtained
by substituting $\hat{a}_{n}$ for $a$. Two common obstacles to employing
the Fisher information in this way are: (1) the failure of $W\sim F_{0}\left(w\right)$
to satisfy the first condition in Footnote 12 because one or both
endpoints of the sample space ($\mathbb{W}$) depend on $a$; and
(2) the non-concavity of the log-likelihood, ${\textstyle \sum_{i=1}^{n}}\ln\left(f_{0}\left(W_{i}\right)\right)$,
as a function of $a$.\newpage{}

\subsection{$\boldsymbol{k^{\textrm{th}}}$ Root of Unnormalized Jeffreys Prior}

\noindent For probability distributions with $k\geq2$ parameters
(i.e., $\boldsymbol{a}=\left[a_{1},\ldots,a_{k}\right]$), the Fisher
information for $n=1$ generalizes to the $k\times k$ Fisher information
matrix,{\small{}
\begin{equation}
\mathbf{I}_{F_{0}}^{\left(k\right)}\left(\boldsymbol{a}\right)=\textrm{E}_{F_{0}}\left[\begin{array}{ccc}
\left(\dfrac{\partial\ln\left(f_{0}\left(W\right)\right)}{\partial a_{1}}\right)^{2} & \cdots & \left(\dfrac{\partial\ln\left(f_{0}\left(W\right)\right)}{\partial a_{1}}\right)\left(\dfrac{\partial\ln\left(f_{0}\left(W\right)\right)}{\partial a_{k}}\right)\\
\vdots & \ddots & \vdots\\
\left(\dfrac{\partial\ln\left(f_{0}\left(W\right)\right)}{\partial a_{k}}\right)\left(\dfrac{\partial\ln\left(f_{0}\left(W\right)\right)}{\partial a_{1}}\right) & \cdots & \left(\dfrac{\partial\ln\left(f_{0}\left(W\right)\right)}{\partial a_{k}}\right)^{2}
\end{array}\right]{\color{red},}
\end{equation}
}which is well defined subject to natural extensions of the regularity
conditions stated in Footnote 12.

Since the determinant of this matrix represents the volume spanned
by the parallelepiped formed by the $k$ columns (or rows), it constitutes
a natural extension to $k$ dimensions of the scalar measure of the
average impact of $a$ on the PMF/PDF in (11). One then can raise
this quantity to the $\tfrac{1}{2k}$ power to: (i) reduce its dimensional
units from $k$ to 1, allowing it to be expressed on a ``per-parameter''
basis; and (ii) return the squared units formed by all matrix elements
in (14) to their original scale (in the same way exponentiating the
Shannon entropy did in Section 3). Interestingly, the resulting quantity,
$\left(\det\left(\mathbf{I}_{F_{0}}^{\left(k\right)}\left(\boldsymbol{a}\right)\right)\right)^{1/2k}$,
is equivalent to the $k^{\textrm{th}}$ root of the unnormalized Jeffreys
prior PMF/PDF for the parameter vector $\boldsymbol{a}=\left[a_{1},a_{2},\ldots,a_{k}\right]$:
\begin{equation}
p_{\textrm{J}}\left(a_{1},\ldots,a_{k}\right)\propto\sqrt{\det\left(\mathbf{I}_{F_{0}}^{\left(k\right)}\left(\boldsymbol{a}\right)\right)}.
\end{equation}

Although $\left(\det\left(\mathbf{I}_{F_{0}}^{\left(k\right)}\left(\boldsymbol{a}\right)\right)\right)^{1/2k}$
appears to be a good candidate for measuring functional adaptability,
a closer examination reveals a significant shortcoming. By rewriting
(14) as
\begin{equation}
\mathbf{I}_{F_{0}}^{\left(k\right)}\left(\boldsymbol{a}\right)=\textrm{E}_{F_{0}}\left[\left(\dfrac{1}{f_{0}\left(W\right)}\right)^{2}\mathbf{D}^{\left(k\right)}\left(\boldsymbol{a}\right)\right],
\end{equation}
with
\[
\mathbf{D}^{\left(k\right)}\left(\boldsymbol{a}\right)=\left[\begin{array}{ccc}
\left(\dfrac{\partial f_{0}\left(W\right)}{\partial a_{1}}\right)^{2} & \cdots & \left(\dfrac{\partial f_{0}\left(W\right)}{\partial a_{1}}\right)\left(\dfrac{\partial f_{0}\left(W\right)}{\partial a_{k}}\right)\\
\vdots & \ddots & \vdots\\
\left(\dfrac{\partial f_{0}\left(W\right)}{\partial a_{k}}\right)\left(\dfrac{\partial f_{0}\left(W\right)}{\partial a_{1}}\right) & \cdots & \left(\dfrac{\partial f_{0}\left(W\right)}{\partial a_{k}}\right)^{2}
\end{array}\right],
\]
one can see that
\begin{equation}
\left(\det\left(\mathbf{I}_{F_{0}}^{\left(k\right)}\left(\boldsymbol{a}\right)\right)\right)^{1/2k}=\left(\det\left(\textrm{E}_{F_{0}}\left[\left(\dfrac{1}{f_{0}\left(W\right)}\right)^{2}\mathbf{D}^{\left(k\right)}\left(\boldsymbol{a}\right)\right]\right)\right)^{1/2k},
\end{equation}
where the component $\left(\tfrac{1}{f_{0}\left(W\right)}\right)^{2}$
is identical to that inside the expectation operation in (10) for
$p=2$. Thus, (17) reflects not only the distribution's adaptability
(through $\mathbf{D}^{\left(k\right)}\left(\boldsymbol{a}\right)$),
but also its simplicity, rendering it inappropriate for a measure
of functional adaptability alone.

Rather fortuitously, this particular drawback actually identifies
(17) as a good candidate for measuring the functional versatility
of $F_{0}$, since such a measure must capture both simplicity and
adaptability at the same time. Of course, the indicated expression
is not unique in this regard. For example, one easily could modify
(17) by replacing $\left(\tfrac{1}{f_{0}\left(W\right)}\right)^{2}$
with $\left|\tfrac{1}{f_{0}\left(W\right)}\right|^{p}$ for any $p\in\mathbb{R}_{0}^{+}$
and then raising the resulting determinant to the $\tfrac{1}{pk}$
power. However, (17) clearly offers the benefit of maximum analytical
familiarity because it is based on the extensively studied Fisher
information matrix.

Given that $\mathbf{I}_{F_{0}}^{\left(k\right)}\left(\boldsymbol{a}\right)$
depends on the parameter vector, $\boldsymbol{a}$, we still need
to average over the prior distribution of $\boldsymbol{a}$ to ensure
that the family of probability distributions specified by the null
hypothesis in (1) is unconditionally versatile (i.e., across all values
of $\boldsymbol{a}$). This yields the tentative measure
\[
\mathcal{V}\left(F_{0}\right)=\left(\det\left(\textrm{E}_{P}\left[\mathbf{I}_{F_{0}}^{\left(k\right)}\left(\boldsymbol{a}\right)\right]\right)\right)^{1/2k}
\]
\begin{equation}
=\left(\det\left(\textrm{E}_{P}\left[\textrm{E}_{F_{0}}\left[\left(\dfrac{1}{f_{0}\left(W\right)}\right)^{2}\mathbf{D}^{\left(k\right)}\left(\boldsymbol{a}\right)\right]\right]\right)\right)^{1/2k},
\end{equation}
where $P\equiv P\left(\boldsymbol{a}\right)$ denotes the prior distribution
function. The new component, $\textrm{E}_{P}\left[\mathbf{I}_{F_{0}}^{\left(k\right)}\left(\boldsymbol{a}\right)\right]$,
is commonly called the Bayesian Fisher information matrix in the relevant
literature.\footnote{See, e.g., Daniels and Hogan (2008).}

In the next section, we refine and formalize the versatility measure
of (18). Then, in Section 6, we illustrate its usefulness in risk
analysis through application to a number of common frequency and severity
models.

\section{Measuring Versatility}

\subsection{Maximum-Entropy Prior}

\noindent The measure of functional versatility stated in (18) remains
incomplete because its indicated prior distribution, $P\equiv P\left(\boldsymbol{a}\right)$,
is unspecified. Here, as in many Bayesian settings with little or
no information about distribution parameters, the choice of an appropriate
prior distribution is somewhat challenging.

In the present context, all we know about the parameter vector $\boldsymbol{a}$
is that it should be representative of all possible vectors employed
to characterize either a frequency or severity distribution. Since
most frequency PMFs and severity PDFs are commonly formulated (or
easily reformulated) with real-valued parameters spanning the space
$\mathbb{R}^{+}$ (as in the Exponential case, where both the mean,
$\theta$, and its inverse, $\lambda$, can be any positive real number),
we will restrict attention to vectors $\boldsymbol{a}$ spanning $\left(\mathbb{R}^{+}\right)^{k}$
and seek a reasonable uninformative prior distribution according to
the principle of indifference. This usually takes one of two approaches:
the method of Jeffreys priors, or that of maximum-entropy priors.\footnote{See Jaynes (2003) for a classic treatment of uninformative prior distributions
and Llorente et al. (2023) for some contemporary perspectives.}

As observed in Subsection 4.2, the Jeffreys prior distribution is
constructed so the joint PMF/PDF is proportional to the determinant
of the square-root of the Fisher information matrix, as shown in (15).
This formulation gives the prior distribution the desirable property
that it is invariant with regard to an invertible and differentiable
transformation of the parameter vector; in other words, if
\[
\left[b_{1},b_{2},\ldots,b_{k}\right]=\boldsymbol{\textrm{g}}\left(a_{1},a_{2},\ldots,a_{k}\right)
\]
for some invertible and differentiable $k$-dimensional vector function
$\boldsymbol{\textrm{g}}\left(\boldsymbol{\cdot}\right)$, then 
\[
p_{\textrm{J}}^{\prime}\left(b_{1},b_{2},\ldots,b_{k}\right)=p_{\textrm{J}}\left(a_{1},a_{2},\ldots,a_{k}\right)\left|\det\left(\boldsymbol{\textrm{J}}\right)\right|,
\]
where
\[
\boldsymbol{\textrm{J}}\equiv\left[\boldsymbol{\textrm{J}}_{i,j}\right]=\left[\dfrac{\partial a_{i}}{\partial b_{j}}\right]
\]
denotes the $k\times k$ Jacobian matrix of the inverse transformation.

Unfortunately, this approach has one major drawback: the Jeffreys
prior distribution often is improper in the sense that its PMF/PDF
sums/integrates to infinity, precluding a well-defined constant of
summation/integration. Sometimes this problem can be finessed if one
is interested only in calculating a posterior distribution (which
may be proper even if the prior is improper). However, it poses an
insurmountable obstacle to evaluating (18) in many cases.\footnote{For example, if $X\sim\textrm{Poisson}\left(\lambda\right)$, then
$p_{J}\left(\lambda\right)=\lambda^{-1/2}$ and the expression in
(18) equals $\left({\textstyle \int_{0}^{\infty}\lambda^{-3/2}d\lambda}\right)^{1/2}=\infty$. }

The maximum-entropy approach, mentioned in Subsections 3.1 and 3.2,
addresses the problem of no prior information by spreading the probability
density as evenly as possible over a given sample space, subject to
certain constraints. This technique always provides a proper PMF/PDF,
but often involves some element of subjectivity in the selection of
hyper-parameters (i.e., the parameters of the maximum-entropy distribution
itself), either as fixed constants or random quantities with their
own hyper-prior distributions. For simplicity, we will assume that
it is reasonable to treat all hyper-parameters as fixed constants
in the present analysis.

Among maximum-entropy distributions defined on $a\in\mathbb{R}^{+}$,
the most obvious possibility is the Exponential distribution, which,
as noted in Section 3, possesses the maximum-entropy PDF subject to
a fixed mean, $\mu$. Unfortunately, the selection of the constant
hyper-parameter $\mu$ poses serious difficulties. This is because
the salient choice \textendash{} $\mu=1$, based on the intuition
that positive parameters can be divided into two mirror realms: real
numbers greater than 1 and their inverses \textendash{} actually divides
$\mathbb{R}^{+}$ unequally, with
\[
\Pr\left\{ a\leq1\right\} =1-e^{-1/\mu}
\]
\[
=1-e^{-1}
\]
\[
\approx0.6321.
\]
Furthermore, all other conspicuous choices for $\mu$ (such as $\mu=\tfrac{1}{\ln\left(2\right)}\approx1.4427$,
which implies a median of 1) appear rather unnatural.

To overcome this problem, we can require the distribution of $\ln\left(a\right)\in\mathbb{R}$
to be symmetric about 0. Here, the maximum-entropy distribution, subject
to a fixed variance, $\sigma^{2}$, is $\textrm{Lognormal}\left(\mu=0,\sigma\right)$,
with $\sigma\in\mathbb{R}^{+}$, and we can set the hyper-parameter
$\sigma$ equal to 1 based on the rationale that $\ln\left(\sigma\right)$,
like $\ln\left(a\right)$, must be symmetric about the origin (which,
for a constant, is exactly 0). Taking the parameters $a_{1},a_{2},\ldots,a_{k}$
in (18) to be IID $\textrm{Lognormal}\left(\mu=0,\sigma=1\right)$
random variables with joint distribution function $P_{\textrm{LN}}\equiv P_{\textrm{LN}}\left(\boldsymbol{a}\right)$
then yields the following expression:
\begin{equation}
\mathcal{V}\left(F_{0}\right)=\left(\det\left(\textrm{E}_{\textrm{LN}}\left[\mathbf{I}_{F_{0}}^{\left(k\right)}\left(\boldsymbol{a}\right)\right]\right)\right)^{1/2k}.
\end{equation}
Since $\textrm{E}_{\textrm{LN}}\left[\left|a_{1}^{p_{1}}a_{2}^{p_{2}}\cdots a_{k}^{p_{k}}\right|\right]<\infty$
for all $\boldsymbol{p}\in\mathbb{R}^{k}$, the right-hand side of
(19) is well defined as long as the individual elements of $\mathbf{I}_{F_{0}}^{\left(k\right)}\left(\boldsymbol{a}\right)$
do not grow faster than $\left|a_{1}^{p_{1}}a_{2}^{p_{2}}\cdots a_{k}^{p_{k}}\right|$
as the individual $a_{i}$ approach $0$ or $\infty$.

As a preliminary assessment of (19) for measuring distributional versatility,
we apply the proposed measure to three 1-parameter distributions \textendash{}
$\textrm{Exponential}\left(\lambda\right)$, $\textrm{Gamma}\left(r,\lambda=1\right)$,
and $\textrm{Weibull}\left(\lambda=1,\tau\right)$ \textendash{} each
of which constitutes a special case of the 3-parameter $\textrm{Generalized Gamma}\left(r,\lambda,\tau\right)$
(GG) family with PDF
\[
f_{Y\mid r,\lambda,\tau}^{\left(\Gamma\right)}\left(y\right)=\dfrac{\tau\lambda^{r}}{\Gamma\left(r\right)}y^{\tau r-1}e^{-\lambda y^{\tau}},
\]
for $r\in\mathbb{R}^{+}$, $\lambda\in\mathbb{R}^{+}$, and $\tau\in\mathbb{R}^{+}$.
For each distribution, the third and fourth columns of Table 2 provide
the mathematical expression for Fisher information and the computed
value of $\mathcal{V}\left(F_{0}\right)$, respectively.\newpage{}
\noindent \begin{center}
Table 2. Versatility Calculations for Three 1-Parameter Continuous
Distributions
\par\end{center}

\noindent \begin{center}
\begin{tabular}{|c|c|c|c|}
\hline 
\textbf{Distribution} & \textbf{PDF} & \textbf{Fisher} & $\boldsymbol{\mathcal{V}\left(F_{0}\right)}$\tabularnewline
\textbf{($\boldsymbol{F_{0}}$)} &  & \textbf{Information} & \tabularnewline
\hline 
\hline 
$\textrm{Exponential}\left(\lambda\right)$ & $f_{Y\mid\lambda}^{\left(\textrm{E}\right)}\left(y\right)=\lambda e^{-\lambda y}$ & $\dfrac{1}{\lambda^{2}}$ & $2.7183$\tabularnewline
$\equiv\textrm{GG}\left(r=1,\lambda,\tau=1\right)$ &  &  & \tabularnewline
\hline 
$\textrm{Gamma}\left(r,\lambda=1\right)$ & $f_{Y\mid r,\lambda=1}^{\left(\Gamma\right)}\left(y\right)=\dfrac{y^{r-1}e^{-y}}{\Gamma\left(r\right)}$ & $\psi^{\left(1\right)}\left(r\right)$ & $2.8399$\tabularnewline
$\equiv\textrm{GG}\left(r,\lambda=1,\tau=1\right)$ &  &  & \tabularnewline
\hline 
$\textrm{Weibull}\left(\lambda=1,\tau\right)$ & $f_{Y\mid\lambda=1,\tau}^{\left(\textrm{Wei}\right)}\left(y\right)=\tau y^{\tau-1}e^{-y^{\tau}}$ & $\dfrac{1}{\tau^{2}}\left[\left(1-\gamma\right)^{2}+\dfrac{\pi^{2}}{6}\right]$ & $3.6709$\tabularnewline
$\equiv\textrm{GG}\left(r=1,\lambda=1,\tau\right)$ &  &  & \tabularnewline
\hline 
\end{tabular}
\par\end{center}

\noindent {\small{}Notes: (i) $\gamma=0.5772156649\ldots$ denotes
Euler's constant; and (ii) $\psi^{\left(1\right)}\left(r\right)=\tfrac{d^{2}\ln\left(\Gamma\left(r\right)\right)}{dr^{2}}$
denotes the polygamma function of order 1.}{\small \par}

\bigskip{}
Examining the indicated values of $\mathcal{V}\left(F_{0}\right)$
in conjunction with representative plots of the corresponding PDFs
in Figure 1 reveals a distinct pattern. Although all three distributions
are relatively flat for most values of $y\in\mathbb{R}_{0}^{+}$,
the versatility measure increases as the distribution is better able
to concentrate probability around its mode.

For $\textrm{Exponential}\left(\lambda\right)$, the mode is always
0 and $f_{Y\mid\lambda}^{\left(\textrm{E}\right)}\left(0\right)=\lambda$
is finite. However, if the modes of $\textrm{Gamma}\left(r,\lambda=1\right)$
and $\textrm{Weibull}\left(\lambda=1,\tau\right)$ are 0 (which occurs
for $r\leq1$ and $\tau\leq1$, respectively), then $\underset{y\rightarrow0^{+}}{\lim}f_{Y\mid r,\lambda=1}^{\left(\Gamma\right)}\left(y\right)=\underset{y\rightarrow0^{+}}{\lim}f_{Y\mid\lambda=1,\tau}^{\left(\textrm{Wei}\right)}\left(y\right)=\infty$
(except for the boundary cases of $r=1$ and $\tau=1$). Although
$\textrm{Gamma}\left(r,\lambda=1\right)$ enjoys the flexibility of
possessing an arbitrarily large mode (given by $r-1$ for $r\geq1$),
the associated value of $f_{Y\mid r,\lambda=1}^{\left(\Gamma\right)}\left(r-1\right)$
is bounded, decreasing from 1 (when $r=1$) to 0 (as $r\rightarrow\infty$).
On the other hand, the mode of $\textrm{Weibull}\left(\lambda=1,\tau\right)$
(given by $\left(\tfrac{\tau-1}{\tau}\right)^{1/\tau}$ for $\tau\geq1$),
which is limited to the interval $\left[0,1\right)$, yields unbounded
values of $f_{Y\mid\lambda=1,\tau}^{\left(\textrm{W}\right)}\left(\left(\tfrac{\tau-1}{\tau}\right)^{1/\tau}\right)$,
increasing from 1 (when $\tau=1$) to $\infty$ (as $\tau\rightarrow\infty$).

Thus, in moving from $\textrm{Exponential}\left(\lambda\right)$ to
$\textrm{Gamma}\left(r,\lambda=1\right)$, the versatility measure
increases because the enhanced adaptability from (i) a higher concentration
of probability at the origin when $r<1$, and (ii) arbitrary placement
of the mode when $r>1$, offsets the attendant decrease in simplicity
from diminished entropy. Likewise, in going from $\textrm{Gamma}\left(r,\lambda=1\right)$
to $\textrm{Weibull}\left(\lambda=1,\tau\right)$, versatility continues
to rise because the greater adaptability from a higher concentration
of probability near $y=1$ (as $\tau\rightarrow\infty$), while retaining
a similar concentration of probability at the origin (when $\tau<1$),
offsets the further decrease in simplicity/entropy.\newpage{}
\noindent \begin{center}
\includegraphics[scale=0.1]{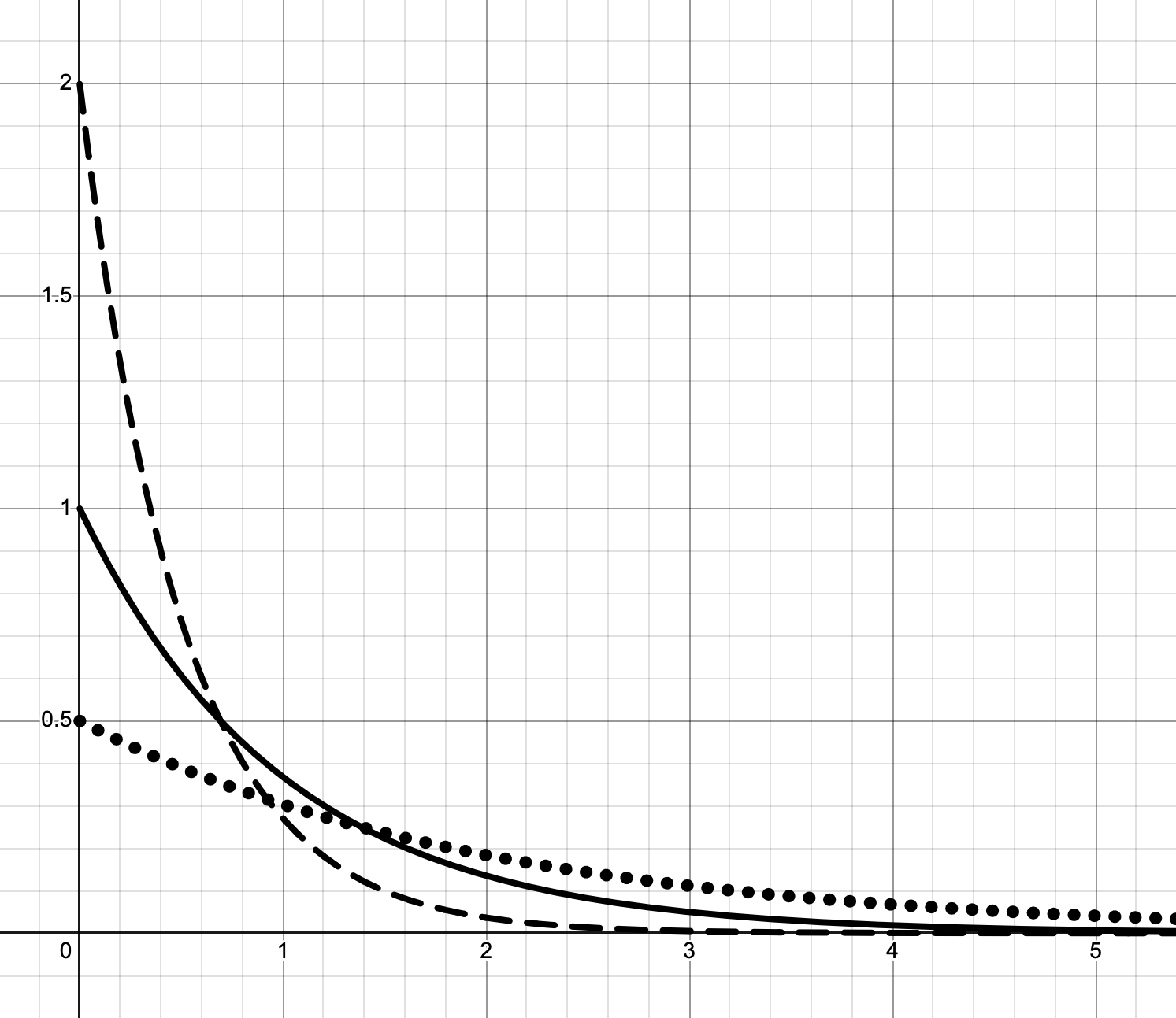}\enskip{}\includegraphics[scale=0.1]{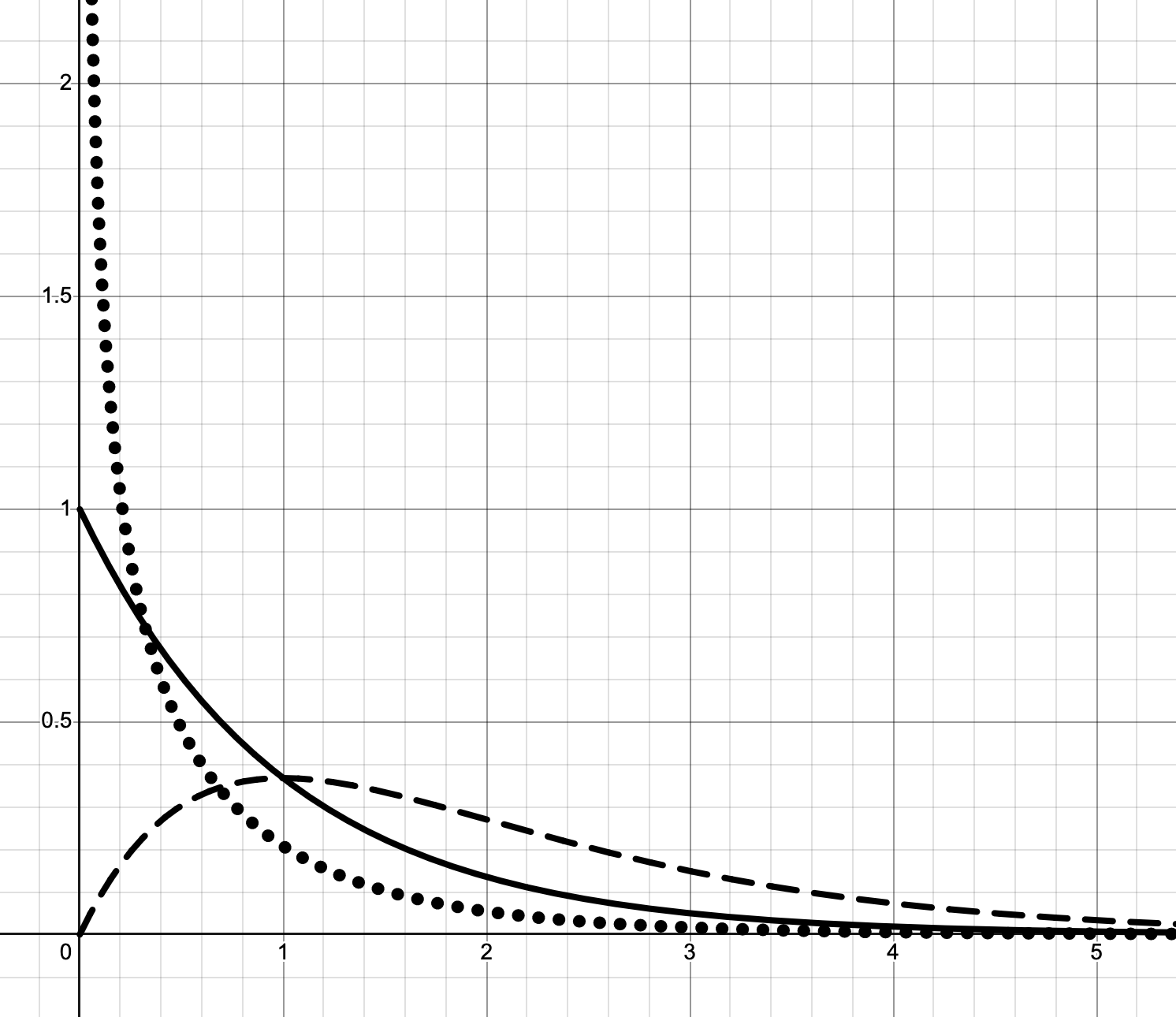}\enskip{}\includegraphics[scale=0.1]{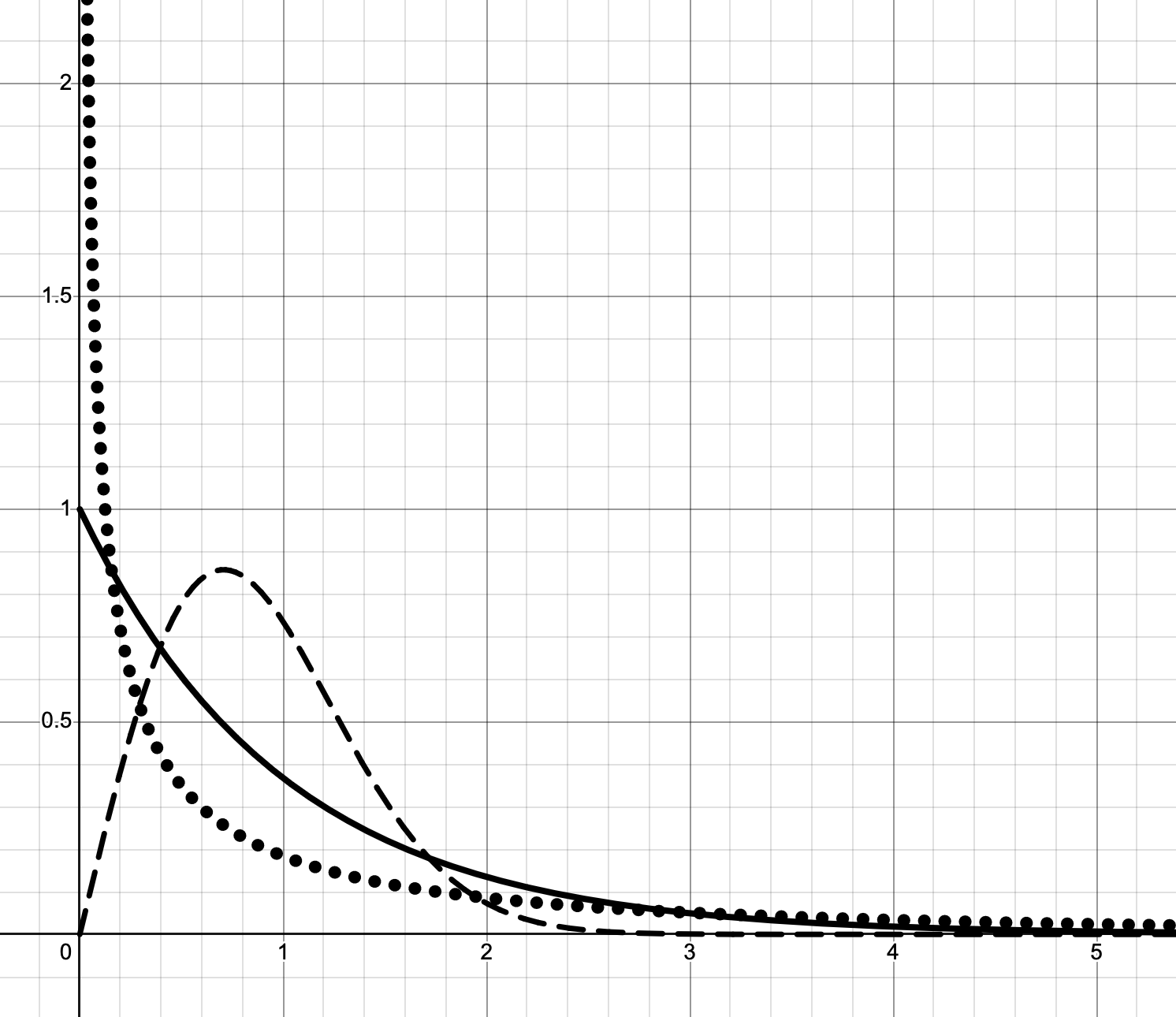}
\par\end{center}

\begin{singlespace}
\noindent \begin{center}
Figure 1. PDFs of the $\textrm{Exponential}\left(\lambda\right)$,
$\textrm{Gamma}\left(r,\lambda=1\right)$, and $\textrm{Weibull}\left(\lambda=1,\tau\right)$
Distributions (from left to right)
\par\end{center}

\noindent \begin{center}
{[}variable parameter $=0.5$ for dotted line, $1.0$ for solid line,
and $2.0$ for dashed line{]}
\par\end{center}
\end{singlespace}

\bigskip{}

Given the definition in (19), it is clear that $\mathcal{V}\left(F_{0}\right)\in\mathbb{R}^{+},$
and thus not bounded above. This makes the interpretation of versatility
magnitudes somewhat challenging, and raises the natural question of
whether it is possible to transpose $\mathcal{V}\left(F_{0}\right)$
to a bounded interval, such as $\left(0,1\right)$, in a way that
affords meaningful interpretations (as in the case of the regression
$R^{2}$, which can be viewed as the proportion of variation explained
by a posited regression model). One salient approach would be to consider
transformations $\textrm{T}\left(\mathcal{V}\left(F_{0}\right)\right):\mathbb{R}^{+}\rightarrow\left(0,1\right)$
that endow a percentile to each possible versatility measure. In other
words, if we could define and identify a theoretical distribution,
$F_{\mathcal{V}}\left(v\right)$, of $\mathcal{V}\left(F_{0}\right)$
over all probability distributions $F_{0}$ within some prescribed
category (say, e.g., severity distributions on $\mathbb{R}_{0}^{+}$),
then we might set $\textrm{T}\left(\mathcal{V}\left(F_{0}\right)\right)=100F_{\mathcal{V}}^{-1}\left(\mathcal{V}\left(F_{0}\right)\right)$.

Although the problem of identifying a suitable theoretical distribution
$F_{\mathcal{V}}\left(v\right)$ appears largely intractable, there
are potential alternatives. For example, one could employ an uninformative
prior on $\mathbb{R}^{+}$ (such as $\textrm{Lognormal}\left(\mu=0,\sigma=1\right)$),
or base the choice of $F_{\mathcal{V}}\left(v\right)$ on empirical,
rather than theoretical, considerations. In fact, the latter approach
is likely to arise naturally in applied work by observing a large
number of versatility measures for a certain category of distributions
$F_{0}$, yielding an empirical distribution of $\mathcal{V}\left(F_{0}\right)$.

For purposes of the current study, we will defer a deeper investigation
of this issue to future research, recognizing that actual computations
of $\mathcal{V}\left(F_{0}\right)$ are useful primarily for comparing
the characteristics of different probability distributions, rather
than assessing the qualities of any single distribution in isolation.
In this way, the proposed versatility measure is comparable to other
unbounded assessment measures, such as Akaike's Information Criterion
and Mallows' $C_{p}$ Statistic.

\subsection{Least-Compressible Parameterization}

\noindent Interestingly, the apparent reasonableness of the $\mathcal{V}\left(F_{0}\right)$
values in Table 2 is easily undermined by reparameterizing the indicated
PDFs, subject to the restriction that each parameter span $\mathbb{R}^{+}$.
For example, consider the two alternative parameterizations of the
Exponential distribution:
\begin{equation}
f_{Y\mid\lambda}^{\left(\textrm{E}\right)}\left(y\right)=\lambda e^{-\lambda y},
\end{equation}
as in Table 2; and

\noindent 
\begin{equation}
f_{Y\mid\beta}^{\left(\textrm{E}\right)}\left(y\right)=\beta^{p}\exp\left(-\beta^{p}y\right),
\end{equation}
where we have replaced $\lambda$ by $\beta^{p}$ for some fixed value
of $p\in\mathbb{R}^{+}$. As a result of this seemingly minor change,
the value of $\mathcal{V}\left(F_{0}\right)$ is multiplied by a factor
of $p$, which means that the versatility measure can be made either
arbitrarily large (as $p\rightarrow\infty$) or arbitrarily small
(as $p\rightarrow0^{+}$).

Such sensitivity to a change in parameters renders the versatility
measure of (19) useless unless we can find a way to invalidate variations
like those in (21) for $p\neq1$. Fortunately, this task is readily
accomplished, at least conceptually, by comparing (21) with the more
conventional parameterization of (20). Since (21) is clearly more
complex (in the Kolmogorov sense) than (20) for any $p\neq1$, we
can remedy the problem of multiple parameterizations by requiring
that only the least-compressible alternative (under some reasonable
coding system for mathematical notation) be used to characterize a
distribution function for measuring its versatility.

As mentioned in the discussion of notational complexity in Subsection
3.1, numerous methods for assessing the complexity of functional forms
have been proposed in the literature, but none is broadly accepted
or highly tractable. However, for the present purpose (which is much
more delimited than the quest for a simplicity measure in the earlier
subsection), it is possible to construct a relatively elementary \textendash{}
albeit informal \textendash{} system for evaluating the compressibility
of PMF/PDF parameterizations for commonly used frequency/severity
distributions.

To this end, we will evaluate PMFs/PDFs according to the following
rules for counting mathematical symbols (and exclude any functions
incompatible with these rules):

$\bullet$\enskip{}Equations must be written from left to right using
the conventional order of common operations:

\qquad{}(1) operations within parentheses;

\qquad{}(2) exponentiation ($\wedge$) and taking roots ($\vee$);

\qquad{}(3) multiplication ($\times$) and division ($/$); and

\qquad{}(4) addition ($+$) and subtraction ($-$).

$\bullet$\enskip{}Each individual real number (positive or negative),
variable, and parameter counts as 1 symbol. 

$\bullet$\enskip{}Each common mathematical operation counts as 1
symbol.

$\bullet$\enskip{}Each parenthesis counts as 1 symbol.

$\bullet$\enskip{}Each elementary function (including the exponential,
standard trigonometric, and hyperbolic functions, along with their
respective inverses) counts as 1 symbol, written as $\phi$ (which
denotes the given function's unique symbol). However, the argument
of any function must be placed in parentheses, which count as 2 symbols.

$\bullet$\enskip{}The gamma function counts as 1 symbol, written
as $\Gamma$.\footnote{Although generally considered a non-elementary function, the gamma
function is explicitly included in our system because of its frequent
use in probability modeling. The beta function, which also is commonly
used, is excluded because it can be written as a simple combination
of gamma functions.}

$\bullet$\enskip{}Other special functions, as well as more complicated
mathematical operations (e.g., summations, limits, derivatives, integrals,
etc.) are not permitted.\footnote{We note that this rule precludes applications to mixture distributions,
which are commonly used in risk analysis. However, two practical considerations
mitigate the impact of this restriction: (1) it remains feasible to
compute and compare the versatility measures of different mixture
distributions whose formulations in terms of integrals/sums over the
mixing distribution are carried out in a consistent manner; and (2)
it is possible to assess the versatility of a mixture distribution
informally by measuring the versatilities of its individual components
(i.e., the kernel and mixing distribution) separately.}

Applying the above system to the PDFs in (20) and (21) yields the
following results:
\begin{equation}
f_{Y\mid\lambda}^{\left(\textrm{E}\right)}\left(y\right)=\lambda\times e\left(\tilde{1}\times\lambda\times y\right)
\end{equation}
(where $e\left(\cdot\right)$ denotes the exponential function and
$\tilde{1}$ denotes $-1$), which requires 10 symbols; and
\begin{equation}
f_{Y\mid\beta}^{\left(\textrm{E}\right)}\left(y\right)=\left(\beta\wedge p\right)\times e\left(\tilde{1}\times\beta\wedge p\times y\right),
\end{equation}
which requires 16 symbols. Thus, according to the proposed evaluation
method, the parameterization given by (20) clearly is more compressible
than that of (21), compelling us to use the former for computing $\mathcal{V}\left(F_{0}\right)$
(unless we can find an even less-compressible alternative).

Naturally, it is possible to write a given PMF/PDF so that it requires
more symbols than necessary. For example, one could re-express (22)
as
\begin{equation}
f_{Y\mid\lambda}^{\left(\textrm{E}\right)}\left(y\right)=\lambda\times\left(\left(e\wedge\left(\tilde{1}\right)\right)\wedge\lambda\right)\wedge y
\end{equation}
and (23) as
\begin{equation}
f_{Y\mid\beta}^{\left(\textrm{E}\right)}\left(y\right)=\left(\beta\wedge p\right)\times\left(\left(e\wedge\left(\tilde{1}\right)\right)\wedge\left(\beta\wedge p\right)\right)\wedge y,
\end{equation}
requiring 15 and 23 symbols, respectively. To avoid this problem,
we assume that all PMFs/PDFs are written as efficiently as possible.

Furthermore, it is possible for two or more distinct parameterizations
to minimize the symbol count under the system outlined above. For
example, setting $\lambda=\tfrac{1}{\theta}$ gives the PDF
\[
f_{Y\mid\theta}^{\left(\textrm{E}\right)}\left(y\right)=\dfrac{1}{\theta}e^{-y/\theta},
\]
which was used in Subsection 3.1. Rewriting this PDF as
\[
f_{Y\mid\theta}^{\left(\textrm{E}\right)}\left(y\right)=e\left(\tilde{1}\times y/\theta\right)/\theta
\]
reveals that it requires 10 symbols, the same number as the PDF of
(20). Although the two corresponding values of $\mathcal{V}\left(F_{0}\right)$
happen to be equal in this case, they sometimes may be different.
If that occurs, then the easiest way to resolve the ambiguity is to
average the indicated versatility measures. (See, e.g., the $\textrm{Negative Binomial}\left(r,m\right)$
and $\textrm{Discrete Weibull}\left(m,\tau\right)$ examples of Subsection
6.2 below.)

Given the preceding discussion of this subsection, we now modify the
functional versatility measure of (19) as follows:
\begin{equation}
\mathcal{V}\left(F_{0}\right)=\left(\det\left(\textrm{E}_{\textrm{LN}}\left[\mathbf{I}_{F_{0}}^{\left(k\right)}\left(\boldsymbol{a^{*}}\right)\right]\right)\right)^{1/2k},
\end{equation}
where $\boldsymbol{a^{*}}$ denotes the parameter vector associated
with the least-compressible parameterization of $F_{W\mid\boldsymbol{a}}^{\left(0\right)}\left(w\right)$.

\section{Illustrative Applications}

\noindent Having defined a formal measure of functional versatility,
we now apply this measure to eight distinct families of 2-parameter
distributions to explore its implications. Four of these families
are defined on $\mathbb{R}_{0}^{+}$, and thus appropriate for modeling
severities, and the other four are defined on $\mathbb{Z}_{0}^{+}$,
and appropriate for frequencies. Following this overview, we will
investigate one of the continuous families more closely to show how
the versatility measure can be employed in practice to select certain
probability distributions over others. 

\subsection{Continuous Distributions}

\noindent The four continuous families \textendash{} $\textrm{Gamma}\left(r,\lambda\right)$,
$\textrm{Weibull}\left(\lambda,\tau\right)$, $\textrm{Pareto 2}\left(\alpha,\vartheta\right)$,
and $\textrm{Lognormal}\left(\ln\left(\nu\right),\sigma\right)$ \textendash{}
are chosen primarily for their common use as severity models. Their
PDFs and versatility measures are presented in Table 3, with all parameters
taken to span $\mathbb{R}^{+}$. To provide some (admittedly limited)
sense of the impact of individual parameters on the versatility measure,
the table includes two 1-parameter special cases of each 2-parameter
family (three of which are identical to the 1-parameter examples shown
in Table 2 of Subsection 5.1).

A brief review of Table 3 yields the following observations:
\begin{itemize}
\item The versatility measures associated with the $\textrm{Pareto 2}\left(\alpha,\vartheta\right)$
family are among the lowest in the table. Given that these PDFs, like
that of the $\textrm{Exponential}\left(\lambda\right)$ distribution,
are relatively flat and strictly decreasing, the low versatility measures
suggest that the simplicity of the PDFs is offset by their limited
adaptability.
\item For three of the four families, the versatility measure of the 2-parameter
distribution falls between the measures of the two 1-parameter special
cases. This property seems intuitive because one would expect the
distributional characteristics contributed by each of the two parameters
individually to be ``averaged'' when taken together. However, it
is important to note that this property does not always hold for extreme
values of the ``fixed'' parameter in the 1-parameter distributions.
For example, as the fixed parameter $\alpha^{*}\rightarrow\infty$
in the 1-parameter $\textrm{Pareto 2}\left(\alpha^{*},\vartheta\right)$,
the versatility measure converges to $2.7183$ (which is greater than
the measure associated with $\textrm{Pareto 2}\left(\alpha,\vartheta\right)$).
Furthermore, as the fixed parameter $\sigma^{*}\rightarrow0$ in the
1-parameter $\textrm{Lognormal}\left(\ln\left(\nu\right),\sigma^{*}\right)$,
the versatility measure diverges to $\infty$ (well above the measure
associated with $\textrm{Lognormal}\left(\ln\left(\nu\right),\sigma\right)$).
Apart from these cautionary observations, the analysis of marginal
contributions to versatility from individual parameters lies beyond
the scope of the present work.
\item The parameters with a direct effect on tail behavior \textendash{}
that is, $\tau$ for the $\textrm{Weibull}\left(\lambda,\tau\right)$
distribution and $\alpha$ for the $\textrm{Pareto 2}\left(\alpha,\vartheta\right)$
distribution \textendash{} appear to have relatively larger effects
on versatility than other parameters.
\item The versatility measure of $\textrm{Lognormal}\left(\ln\left(\nu\right),\sigma=1\right)$
is considerably lower than other measures in the table. This is because
the PDF is quite flat for $\nu>1$, with the function's simplicity
offset by its limited adaptability, and quite concentrated (in the
neighborhood of 0) for $\nu<1$, with the function's adaptability
offset by its inadequate simplicity. Although we do not propose a
specific minimum value of the versatility measure for model selection,
it appears that $\mathcal{V}\left(F_{Y\mid\ln\left(\nu\right),\sigma=1}^{\left(\textrm{LN}\right)}\right)=1.0000$
is a strong indicator of idiosyncrasy in this case. 
\end{itemize}
\begin{center}
Table 3. Versatility Calculations for Four 2-Parameter Continuous
Distributions and Special Cases
\par\end{center}

\noindent \begin{center}
\begin{tabular}{|c|c|c|}
\hline 
\textbf{Distribution} & \textbf{PDF} & $\boldsymbol{\mathcal{V}\left(F_{0}\right)}$\tabularnewline
\textbf{($\boldsymbol{F_{0}}$)} &  & \tabularnewline
\hline 
\hline 
$\left(1\right)\textrm{ Gamma}\left(r,\lambda\right)$ & $f_{Y\mid r,\lambda}^{\left(\Gamma\right)}\left(y\right)=\dfrac{\lambda^{r}y^{r-1}e^{-\lambda y}}{\Gamma\left(r\right)}$ & $3.1264$\tabularnewline
$\textrm{Gamma}\left(r=1,\lambda\right)\equiv\textrm{Exponential}\left(\lambda\right)$ & $f_{Y\mid r=1,\lambda}^{\left(\Gamma\right)}\left(y\right)=\lambda e^{-\lambda y}$ & $2.7183$\tabularnewline
$\textrm{Gamma}\left(r,\lambda=1\right)$ & $f_{Y\mid r,\lambda=1}^{\left(\Gamma\right)}\left(y\right)=\dfrac{y^{r-1}e^{-y}}{\Gamma\left(r\right)}$ & $2.8399$\tabularnewline
\hline 
$\left(2\right)\textrm{ Weibull}\left(\lambda,\tau\right)$ & $f_{Y\mid\lambda,\tau}^{\left(\textrm{Wei}\right)}\left(y\right)=\tau\lambda y^{\tau-1}e^{-\lambda y^{\tau}}$ & $3.4349$\tabularnewline
$\textrm{Weibull}\left(\lambda,\tau=1\right)\equiv\textrm{Exponential}\left(\lambda\right)$ & $f_{Y\mid\lambda,\tau=1}^{\left(\textrm{Wei}\right)}\left(y\right)=\lambda e^{-\lambda y}$ & $2.7183$\tabularnewline
$\textrm{Weibull}\left(\lambda=1,\tau\right)$ & $f_{Y\mid\lambda=1,\tau}^{\left(\textrm{Wei}\right)}\left(y\right)=\tau y^{\tau-1}e^{-y^{\tau}}$ & $3.6709$\tabularnewline
\hline 
$\left(3\right)\textrm{ Pareto 2}\left(\alpha,\vartheta\right)$ & $f_{Y\mid\alpha,\vartheta}^{\left(\textrm{P2}\right)}\left(y\right)=\dfrac{\alpha\vartheta^{\alpha}}{\left(y+\vartheta\right)^{\alpha+1}}$ & $2.0874$\tabularnewline
$\textrm{Pareto 2}\left(\alpha=1,\vartheta\right)$ & $f_{Y\mid\alpha=1,\vartheta}^{\left(\textrm{P2}\right)}\left(y\right)=\dfrac{\vartheta}{\left(y+\vartheta\right)^{2}}$ & $1.5694$\tabularnewline
$\textrm{Pareto 2}\left(\alpha,\vartheta=1\right)$ & $f_{Y\mid\alpha,\vartheta=1}^{\left(\textrm{P2}\right)}\left(y\right)=\dfrac{\alpha}{\left(y+1\right)^{\alpha+1}}$ & $2.7183$\tabularnewline
\hline 
$\left(4\right)\textrm{ Lognormal}\left(\ln\left(\nu\right),\sigma\right)$ & $f_{Y\mid\ln\left(\nu\right),\sigma}^{\left(\textrm{LN}\right)}\left(y\right)=\dfrac{\exp\left(-\tfrac{\left(\ln\left(y\right)-\ln\left(\nu\right)\right)^{2}}{2\sigma^{2}}\right)}{\sqrt{2\pi}\sigma y}$ & $3.2327$\tabularnewline
$\textrm{Lognormal}\left(\ln\left(\nu\right)=0,\sigma\right)$ & $f_{Y\mid\ln\left(\nu\right)=0,\sigma}^{\left(\textrm{LN}\right)}\left(y\right)=\dfrac{\exp\left(-\tfrac{\left(\ln\left(y\right)\right)^{2}}{2\sigma^{2}}\right)}{\sqrt{2\pi}\sigma y}$ & $3.8440$\tabularnewline
$\textrm{Lognormal}\left(\ln\left(\nu\right),\sigma=1\right)$ & $f_{Y\mid\ln\left(\nu\right),\sigma=1}^{\left(\textrm{LN}\right)}\left(y\right)=\dfrac{\exp\left(-\tfrac{\left(\ln\left(y\right)-\ln\left(\nu\right)\right)^{2}}{2}\right)}{\sqrt{2\pi}y}$ & $1.0000$\tabularnewline
\hline 
\end{tabular}\bigskip{}
\par\end{center}

\subsection{Discrete Distributions}

\noindent Table 4 presents the PMFs and versatility measures of four
discrete families \textendash{} $\textrm{Negative Binomial}\left(r,\tfrac{m}{m+1}\right)$,
$\textrm{Discrete Weibull}\left(\tfrac{m}{m+1},\tau\right)$, $\textrm{Waring}\left(\alpha,\vartheta\right)$,
and $\textrm{Generalized Poisson}\left(\lambda,\varsigma\right)$
\textendash{} with all parameters spanning $\mathbb{R}^{+}$. The
first three of these families serve as respective analogues of the
first three continuous families discussed above. In particular:

(1) $\textrm{Negative Binomial}\left(r,\tfrac{m}{m+1}\right)$ models
the sum of IID inter-arrival times from a Bernoulli process in the
same manner as $\textrm{Gamma}\left(r,\lambda\right)$ models the
sum of IID inter-arrival times from a Poisson process;

(2) $\textrm{Discrete Weibull}\left(\tfrac{m}{m+1},\tau\right)$\footnote{See Nakagawa and Osaki (1975).}
possesses a tail parameter, $\tau$, with a functional role similar
to that of the corresponding $\textrm{Weibull}\left(\lambda,\tau\right)$
parameter, and also shares the memoryless $\textrm{Geometric}\left(\tfrac{m}{m+1}\right)$
distribution as a special case with $\textrm{Negative Binomial}\left(r,\tfrac{m}{m+1}\right)$
in the same way $\textrm{Weibull}\left(\lambda,\tau\right)$ shares
the memoryless $\textrm{Exponential}\left(\lambda\right)$ distribution
with $\textrm{Gamma}\left(r,\lambda\right)$; and

(3) $\textrm{Waring}\left(\alpha,\vartheta\right)$ is characterized
by strictly decreasing PMFs with heavy tails (such that for all real
$\kappa\leq\alpha$, $\textrm{E}_{X\mid\alpha,\vartheta}^{\left(\textrm{War}\right)}\left[X^{\kappa}\right]=\infty$),
just as $\textrm{Pareto 2}\left(\alpha,\vartheta\right)$ is characterized
by strictly decreasing PDFs with heavy tails (such that for all real
$\kappa\leq\alpha$, $\textrm{E}_{Y\mid\alpha,\vartheta}^{\left(\textrm{P2}\right)}\left[Y^{\kappa}\right]=\infty$).

Although the fourth discrete family, $\textrm{Generalized Poisson}\left(\lambda,\varsigma\right)$,
is not directly analogous to\linebreak{}
$\textrm{Lognormal}\left(\ln\left(\nu\right),\sigma\right)$, both
of these families often serve as well-known benchmarks for the study
of other 2-parameter models.

From Table 4, we make the following observations:
\begin{itemize}
\item For the $\textrm{Negative Binomial}\left(r,\tfrac{m}{m+1}\right)$
family with $\textrm{E}_{X\mid r,\tfrac{m}{m+1}}\left[X\right]=\tfrac{1}{m}$,
there is a second, equally compressible parameterization, $\textrm{Negative Binomial}\left(r,\tfrac{1}{m+1}\right)$,
for which $\textrm{E}_{X\mid r,\tfrac{1}{m+1}}\left[X\right]=m$.
This is analogous to the fact that $\textrm{Gamma}\left(r,\lambda\right)$
possesses the equally compressible counterpart $\textrm{Gamma}\left(r,\theta\right)$
in the continuous case (as noted in Subsection 5.2). However, unlike
the continuous case, the two different parameterizations yield different
versatility measures, which we resolve by averaging (as suggested
in Subsection 5.2). The same issue arises for the $\textrm{Discrete Weibull}\left(\tfrac{m}{m+1},\tau\right)$
family, and is handled in the same way. The versatility measures associated
with the first three discrete families are consistently lower than
the corresponding measures of their continuous analogues. This arises
from the same phenomenon mentioned in Footnote 20: the discrete sample
space $x\in\mathbb{Z}_{0}^{+}$ affords dramatically less opportunity
for functional adaptability in the neighborhood of $0$ than does
the continuous sample space $y\in\mathbb{R}_{0}^{+}$.
\item For all four families, the versatility measure of the 2-parameter
distribution falls between the measures of the two 1-parameter special
cases. However, as noted with regard to Table 3, this property does
not always hold for extreme values of the fixed parameter in the 1-parameter
distributions.
\item The parameters with a direct effect on tail behavior \textendash{}
that is, $\tau$ for the averaged $\textrm{Discrete}$\linebreak{}
$\textrm{Weibull}\left(\tfrac{m}{m+1},\tau\right)$ and $\textrm{Discrete Weibull}\left(\tfrac{1}{m+1},\tau\right)$
distributions and $\alpha$ for the $\textrm{Waring}\left(\alpha,\vartheta\right)$
distribution \textendash{} appear to have relatively larger effects
on versatility than other parameters.\footnote{The versatility measures of distributions parameterized by $\tfrac{1}{m+1}$
are noticeably smaller than those of distributions based on $\tfrac{m}{m+1}$
because the former PMFs are particularly sensitive to $m$ for values
of $x$ close to $0$, whereas the latter are relatively more sensitive
as $x\rightarrow\infty$. For the discrete sample space $x\in\mathbb{Z}_{0}^{+}$,
there is little opportunity for a PMF to benefit from functional adaptability
in the neighborhood of $0$, hence less functional versatility.}
\end{itemize}
\newpage{}
\noindent \begin{center}
Table 4. Versatility Calculations for Four 2-Parameter Discrete Distributions
and Special Cases
\par\end{center}

\noindent \begin{center}
\begin{tabular}{|c|c|c|}
\hline 
\textbf{Distribution} & \textbf{PMF} & $\boldsymbol{\mathcal{V}\left(F_{0}\right)}$\tabularnewline
\textbf{($\boldsymbol{F_{0}}$)} &  & \tabularnewline
\hline 
\hline 
$\left(1\right)\textrm{ N. Bin.}\left(r,\tfrac{m}{m+1}\right)$ & $f_{X\mid r,\tfrac{m}{m+1}}^{\left(\textrm{NB}\right)}\left(x\right)=\dfrac{\Gamma\left(x+r\right)}{\Gamma\left(r\right)\Gamma\left(x+1\right)}\left(\dfrac{m}{m+1}\right)^{r}\left(\dfrac{1}{m+1}\right)^{x}$ & ($1.7910$)\tabularnewline
$\textrm{N. Bin.}\left(r,\tfrac{1}{m+1}\right)$ & $f_{X\mid r,\tfrac{1}{m+1}}^{\left(\textrm{NB}\right)}\left(x\right)=\dfrac{\Gamma\left(x+r\right)}{\Gamma\left(r\right)\Gamma\left(x+1\right)}\left(\dfrac{1}{m+1}\right)^{r}\left(\dfrac{m}{m+1}\right)^{x}$ & ($1.1721$)\tabularnewline
 & Average & $1.4816$\tabularnewline
$\textrm{N. Bin.}\left(r=1,\tfrac{m}{m+1}\right)\equiv\textrm{Geom.}\left(\tfrac{m}{m+1}\right)$ & $f_{X\mid r,\tfrac{m}{m+1}}^{\left(\textrm{NB}\right)}\left(x\right)=\left(\dfrac{m}{m+1}\right)\left(\dfrac{1}{m+1}\right)^{x}$ & ($2.4981$)\tabularnewline
$\textrm{N. Bin.}\left(r=1,\tfrac{1}{m+1}\right)\equiv\textrm{Geom.}\left(\tfrac{1}{m+1}\right)$ & $f_{X\mid r,\tfrac{1}{m+1}}^{\left(\textrm{NB}\right)}\left(x\right)=\left(\dfrac{1}{m+1}\right)\left(\dfrac{m}{m+1}\right)^{x}$ & ($1.0718$)\tabularnewline
 & Average & $1.7850$\tabularnewline
$\textrm{N. Bin.}\left(r,\tfrac{m}{m+1}=\tfrac{1}{m+1}=\tfrac{1}{2}\right)$ & $f_{X\mid r,\tfrac{m}{m+1}=\tfrac{1}{m+1}=\tfrac{1}{2}}^{\left(\textrm{NB}\right)}\left(x\right)=\dfrac{\Gamma\left(x+r\right)}{\Gamma\left(r\right)\Gamma\left(x+1\right)}\left(\dfrac{1}{2}\right)^{x+r}$ & $1.0151$\tabularnewline
\hline 
$\left(2\right)\textrm{ D. Wei.}\left(\tfrac{m}{m+1},\tau\right)$ & $f_{X\mid\tfrac{m}{m+1},\tau}^{\left(\textrm{DWei}\right)}\left(x\right)=\left(\dfrac{1}{m+1}\right)^{x^{\tau}}-\left(\dfrac{1}{m+1}\right)^{\left(x+1\right)^{\tau}}$ & ($2.4379$)\tabularnewline
$\textrm{D. Wei.}\left(\tfrac{1}{m+1},\tau\right)$ & $f_{X\mid\tfrac{1}{m+1},\tau}^{\left(\textrm{DWei}\right)}\left(x\right)=\left(\dfrac{m}{m+1}\right)^{x^{\tau}}-\left(\dfrac{m}{m+1}\right)^{\left(x+1\right)^{\tau}}$ & ($1.6791$)\tabularnewline
 & Average & $2.0585$\tabularnewline
$\textrm{D. Wei.}\left(\tfrac{m}{m+1},\tau=1\right)\equiv\textrm{Geom.}\left(\tfrac{m}{m+1}\right)$ & $f_{X\mid\tfrac{m}{m+1},\tau=1}^{\left(\textrm{DWei}\right)}\left(x\right)=\left(\dfrac{m}{m+1}\right)\left(\dfrac{1}{m+1}\right)^{x}$ & ($2.4981$)\tabularnewline
$\textrm{D. Wei.}\left(\tfrac{1}{m+1},\tau=1\right)\equiv\textrm{Geom.}\left(\tfrac{1}{m+1}\right)$ & $f_{X\mid\tfrac{1}{m+1},\tau=1}^{\left(\textrm{DWei}\right)}\left(x\right)=\left(\dfrac{1}{m+1}\right)\left(\dfrac{m}{m+1}\right)^{x}$ & ($1.0718$)\tabularnewline
 & Average & $1.7850$\tabularnewline
$\textrm{D. Wei.}\left(\tfrac{m}{m+1}=\tfrac{1}{m+1}=\tfrac{1}{2},\tau\right)$ & $f_{X\mid\tfrac{m}{m+1}=\tfrac{1}{m+1}=\tfrac{1}{2},\tau}^{\left(\textrm{DWei}\right)}\left(x\right)=\left(\dfrac{1}{2}\right)^{x^{\tau}}-\left(\dfrac{1}{2}\right)^{\left(x+1\right)^{\tau}}$ & $2.7160$\tabularnewline
\hline 
$\left(3\right)\textrm{ Waring}\left(\alpha,\vartheta\right)$ & $f_{X\mid\alpha,\vartheta}^{\left(\textrm{War}\right)}\left(x\right)=\dfrac{\alpha\Gamma\left(\alpha+\vartheta\right)\Gamma\left(x+\vartheta\right)}{\Gamma\left(\vartheta\right)\Gamma\left(x+\alpha+\vartheta+1\right)}$ & $1.3997$\tabularnewline
$\textrm{Waring}\left(\alpha=1,\vartheta\right)$ & $f_{X\mid\alpha=1,\vartheta}^{\left(\textrm{War}\right)}\left(x\right)=\dfrac{\vartheta}{\left(x+\vartheta\right)\left(x+\vartheta+1\right)}$ & $0.9423$\tabularnewline
$\textrm{Waring}\left(\alpha,\vartheta=1\right)$ & $f_{X\mid\alpha,\vartheta=1}^{\left(\textrm{War}\right)}\left(x\right)=\dfrac{\alpha\Gamma\left(\alpha+1\right)\Gamma\left(x+1\right)}{\Gamma\left(x+\alpha+2\right)}$ & $2.2441$\tabularnewline
\hline 
$\left(4\right)\textrm{ Gen. Poisson}\left(\lambda,\varsigma\right)$ & $f_{X\mid\lambda,\varsigma}^{\left(\textrm{GP}\right)}\left(x\right)=\dfrac{\lambda e^{-\left(\varsigma x+\lambda\right)}\left(\varsigma x+\lambda\right)^{x-1}}{\Gamma\left(x+1\right)}$ & $1.3794$\tabularnewline
$\textrm{Gen. Poisson}\left(\lambda=1,\varsigma\right)$ & $f_{X\mid\lambda=1,\varsigma}^{\left(\textrm{GP}\right)}\left(x\right)=\dfrac{e^{-\left(\varsigma x+1\right)}\left(\varsigma x+1\right)^{x-1}}{\Gamma\left(x+1\right)}$ & $1.7124$\tabularnewline
$\textrm{Gen. Poisson}\left(\lambda,\varsigma=0\right)\equiv\textrm{Poisson}\left(\lambda\right)$ & $f_{X\mid\lambda,\varsigma=0}^{\left(\textrm{GP}\right)}\left(x\right)=\dfrac{e^{-\lambda}\lambda^{x}}{\Gamma\left(x+1\right)}$ & $1.2840$\tabularnewline
\hline 
\end{tabular}
\par\end{center}

\newpage{}
\begin{itemize}
\item The versatility measures of $\textrm{Negative Binomial}\left(r,\tfrac{m}{m+1}=\tfrac{1}{2}\right)$
and $\textrm{Waring}\left(\alpha=1,\vartheta\right)$ are noticeably
lower than other measures in the table. This is because the two PMFs
are quite flat for all values of $r$ and $\vartheta$, respectively,
with both functions' simplicity offset by limited adaptability. As
noted in conjunction with $\textrm{Lognormal}\left(\ln\left(\nu\right),\sigma=1\right)$
in the previous subsection, we do not propose a specific minimum cutoff
for the versatility measure; however, it appears that both $\mathcal{V}\left(F_{X\mid r,\tfrac{m}{m+1}=\tfrac{1}{2}}^{\left(\textrm{NB}\right)}\right)=1.0151$
and $\mathcal{V}\left(F_{X\mid\alpha=1,\vartheta}^{\left(\textrm{War}\right)}\right)=0.9423$
are strong indicators of idiosyncrasy. 
\end{itemize}

\subsection{Pareto 2$\left(\alpha,\vartheta\right)$ vs. Inverse Gamma$\left(r,\lambda\right)$}

\noindent In Subsection 6.1, we noted that the 2-parameter $\textrm{Pareto 2}\left(\alpha,\vartheta\right)$
family possesses a rather low versatility measure compared to the
other 2-parameter continuous distributions in Table 3. This observation
affords the opportunity to consider whether, despite its common use,
$\textrm{Pareto 2}\left(\alpha,\vartheta\right)$ is unduly idiosyncratic.
Given the distribution's obvious simplicity, the basic question is
whether it is insufficiently adaptable for modeling severities effectively,
and therefore should be replaced by other distributions with similar
characteristics (in particular, heavy tails) but greater adaptability.

One potential alternative is the $\textrm{Inverse Gamma}\left(r,\lambda\right)$
distribution, associated with the random variable $Y$ such that $\dfrac{1}{Y}\sim\textrm{Gamma}\left(r,\lambda\right)$.
The $\textrm{Inverse Gamma}\left(r,\lambda\right)$ PDF is given by
\[
f_{Y\mid r,\lambda}^{\left(\textrm{I}\Gamma\right)}\left(y\right)=\dfrac{\lambda^{r}e^{-\lambda/y}}{y^{r+1}\Gamma\left(r\right)},
\]
where the tail parameter, $r$, plays a role comparable to the $\textrm{Pareto 2}\left(\alpha,\vartheta\right)$'s
$\alpha$ in that for all real $\kappa\leq r$, $\textrm{E}_{Y\mid r,\lambda}^{\left(\textrm{I}\Gamma\right)}\left[Y^{\kappa}\right]=\infty$.
As illustrated in Figure 2, the PDFs of the two distributions can
differ considerably near the lower end of the sample space, with $\textrm{Pareto 2}\left(\alpha,\vartheta\right)$
possessing a mode at 0 (where $f_{Y\mid\alpha,\vartheta}^{\left(\textrm{P2}\right)}\left(0\right)=\tfrac{\alpha}{\vartheta}>0$)
and $\textrm{Inverse Gamma}\left(r,\lambda\right)$ increasing from
$f_{Y\mid r,\lambda}^{\left(\textrm{I}\Gamma\right)}\left(0\right)=0$
until it reaches an internal mode at $\tfrac{\lambda}{r+1}$. This
greater flexibility, and therefore adaptability, of $\textrm{Inverse Gamma}\left(r,\lambda\right)$
is reflected in its versatility measure of $\mathcal{V}\left(F_{Y\mid r,\lambda}^{\left(\textrm{I}\Gamma\right)}\right)=3.1264$,
which is equal to that of the $\textrm{Gamma}\left(r,\lambda\right)$
distribution and substantially greater than $\mathcal{V}\left(F_{Y\mid\alpha,\vartheta}^{\left(\textrm{P2}\right)}\right)=2.0874$.
\noindent \begin{center}
\includegraphics[scale=0.1]{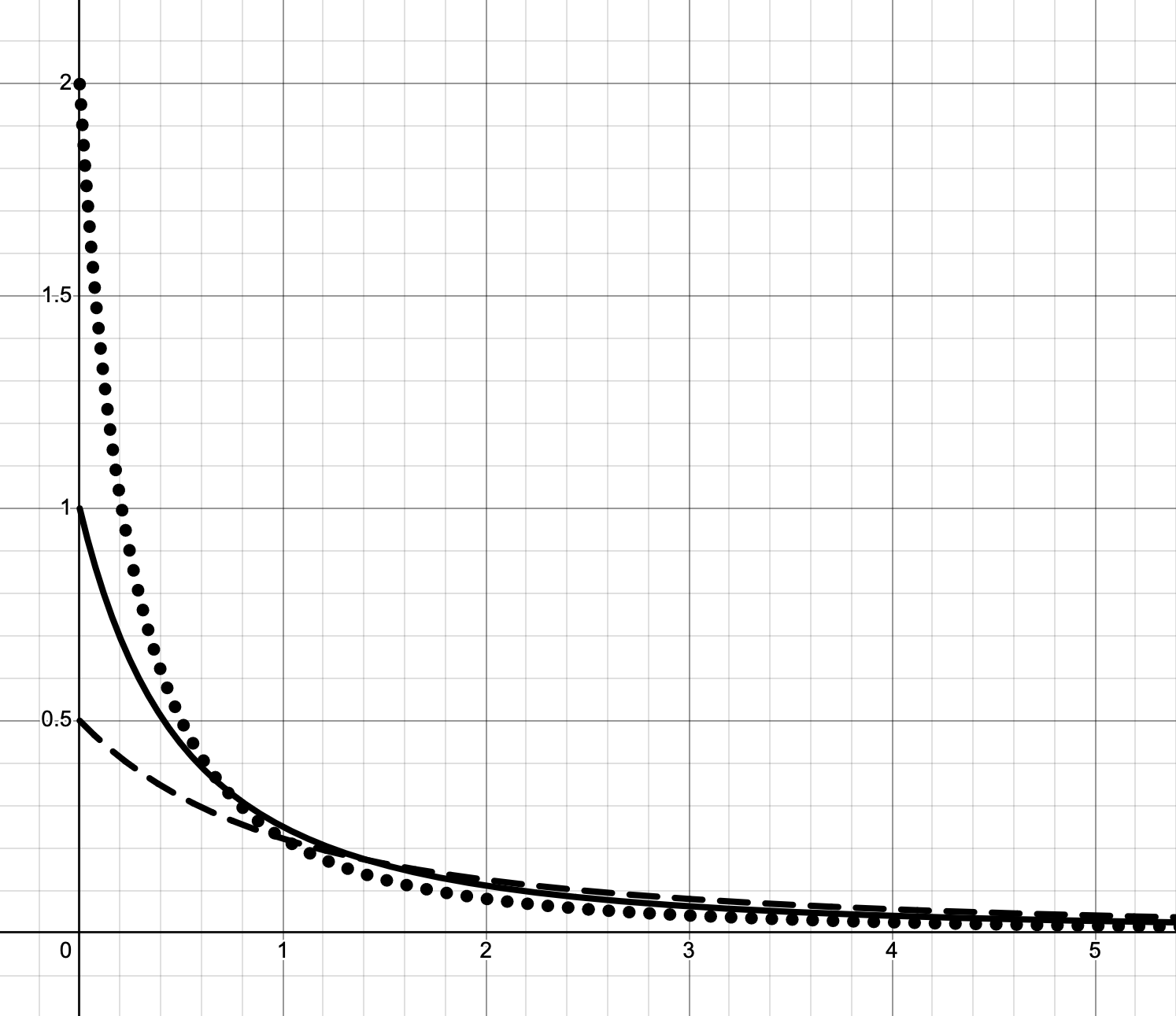}\qquad{}\includegraphics[scale=0.1]{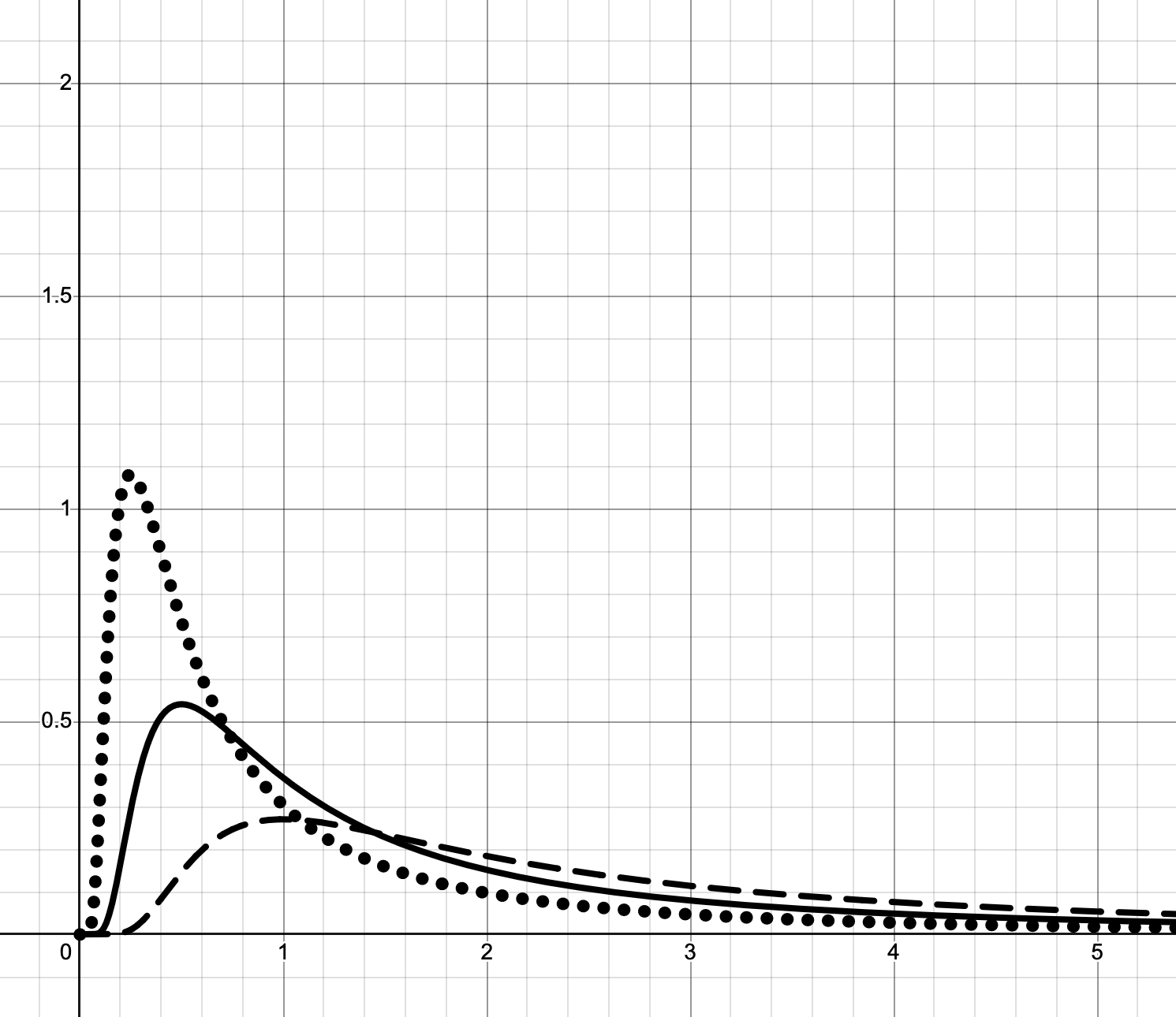}\enskip{}
\par\end{center}

\begin{singlespace}
\noindent \begin{center}
Figure 2. PDFs of $\textrm{Pareto 2}\left(\alpha=1,\vartheta\right)$
and $\textrm{Inverse Gamma}\left(r=1,\lambda\right)$ Distributions
(from left to right)
\par\end{center}

\noindent \begin{center}
{[}variable parameter $=0.5$ for dotted line, $1.0$ for solid line,
and $2.0$ for dashed line{]}
\par\end{center}
\end{singlespace}

\medskip{}

To investigate the relative effectiveness of the $\textrm{Pareto 2}\left(\alpha,\vartheta\right)$
and $\textrm{Inverse Gamma}\left(r,\lambda\right)$ models in practice,
we perform a series of four simple simulation experiments (I, II,
III, and IV) to evaluate each model's ability to distinguish itself
from the other model within a hypothesis-testing framework:
\begin{itemize}
\item In Experiment I, we take $\textrm{Pareto 2}\left(\alpha,\vartheta\right)$
as the null hypothesis, generate $10,000$ IID pairs\linebreak{}
$\left(\alpha_{i}\sim\textrm{Lognormal}\left(\mu=0,\sigma=1\right),\vartheta_{i}\sim\textrm{Lognormal}\left(\mu=0,\sigma=1\right)\right)$,
and for each pair generate $n$ IID observations $Y_{j}\sim\textrm{Pareto 2}\left(\alpha_{i},\vartheta_{i}\right)$
(where $n\in\left\{ 50,100,500,1000,5000,\ldots,100,000\right\} $).
We then use the $n$ IID $Y_{j}$ to perform a minimum chi-squared
goodness-of-fit test (at the $0.05$ level of significance),\footnote{In all minimum chi-squared tests, we employ maximum likelihood to
estimate parameters and Sturges' rule to set the number of test bins.} and record the proportion of false rejections (out of the 10,000
$\left(\alpha_{i},\vartheta_{i}\right)$ pairs) as an estimate of
$\Pr\left\{ \textrm{Type 1 Error}\right\} $.
\item In Experiment II, we take $\textrm{Pareto 2}\left(\alpha,\vartheta\right)$
as the null hypothesis, generate $10,000$ IID pairs\linebreak{}
$\left(r_{i}\sim\textrm{Lognormal}\left(\mu=0,\sigma=1\right),\lambda_{i}\sim\textrm{Lognormal}\left(\mu=0,\sigma=1\right)\right)$,
and for each pair generate $n$ IID observations $Y_{j}\sim\textrm{Inverse Gamma}\left(r_{i},\lambda_{i}\right)$
(where $n\in\left\{ 50,100,500,1000,5000,\ldots,100,000\right\} $).
We then use the $n$ IID $Y_{j}$ to perform a minimum chi-squared
goodness-of-fit test (at the $0.05$ level), and record the proportion
of false retentions (out of the 10,000 $\left(r_{i},\lambda_{i}\right)$
pairs) as an estimate of $\Pr\left\{ \textrm{Type 2 Error}\right\} $.
\item Experiments III and IV are analogous to Experiments I and II, respectively,
with the roles of\linebreak{}
$\textrm{Pareto 2}\left(\alpha,\vartheta\right)$ and $\textrm{Inverse Gamma}\left(r,\lambda\right)$
reversed.
\end{itemize}
\qquad{}The experimental results in Table 5 show rather clearly that
$\textrm{Inverse Gamma}\left(r,\lambda\right)$ performs better on
the whole than $\textrm{Pareto 2}\left(\alpha,\vartheta\right)$.
Whereas both models yield satisfactorily low frequencies of Type 1
error (approaching the $0.05$ level of significance from below as
$n$ increases), the $\textrm{Pareto 2}\left(\alpha,\vartheta\right)$
hypothesis is consistently poor in recognizing and rejecting $\textrm{Inverse Gamma}\left(r,\lambda\right)$
observations (with Type 2 error rates above $0.80$). This shortcoming,
which does not improve for larger sample sizes, appears to derive
from the $\textrm{Pareto 2}\left(\alpha,\vartheta\right)$'s insufficient
sensitivity to variations in the frequency of observations at the
lower end of the sample space. In short, the model gives too much
credence to observations consistent with the hypothesized tail behavior,
thereby failing to discern the presence of otherwise non-$\textrm{Pareto 2}\left(\alpha,\vartheta\right)$
sample characteristics. Although too limited to be conclusive, the
present analysis provides strong prima facie support for using the
proposed versatility measure to identify and eschew the $\textrm{Pareto 2}\left(\alpha,\vartheta\right)$
model's idiosyncrasy.

\medskip{}
\noindent \begin{center}
Table 5. Hypothesis-Test Results from Simulated Data
\par\end{center}

\noindent \begin{center}
\begin{tabular}{|c|c|c|c|c|}
\hline 
 & \textbf{\small{}I. $\boldsymbol{H_{0}}$: Pareto 2} & \textbf{\small{}II. $\boldsymbol{H_{0}}$: Pareto 2} & \textbf{\small{}III. $\boldsymbol{H_{0}}$: Inverse Gamma} & \textbf{\small{}IV. $\boldsymbol{H_{0}}$: Inverse Gamma}\tabularnewline
 & \textbf{\small{}(Pareto 2 Obs.)} & \textbf{\small{}(Inverse Gamma Obs.)} & \textbf{\small{}(Inverse Gamma Obs.)} & \textbf{\small{}(Pareto 2 Obs.)}\tabularnewline
\hline 
\textbf{\small{}Sample Size (}\textbf{\emph{\small{}n}}\textbf{\small{})} & \textbf{\small{}Pr\{Type 1 Error\}} & \textbf{\small{}Pr\{Type 2 Error\}} & \textbf{\small{}Pr\{Type 1 Error\}} & \textbf{\small{}Pr\{Type 2 Error\}}\tabularnewline
\hline 
\hline 
\textbf{\small{}$50$} & \textbf{\small{}$0.0000$} & \textbf{\small{}$1.0000$} & \textbf{\small{}$0.0034$} & \textbf{\small{}$1.0000$}\tabularnewline
\hline 
\textbf{\small{}$100$} & \textbf{\small{}$0.0056$} & \textbf{\small{}$0.8671$} & \textbf{\small{}$0.0131$} & \textbf{\small{}$0.9999$}\tabularnewline
\hline 
\textbf{\small{}$500$} & \textbf{\small{}$0.0181$} & \textbf{\small{}$0.8113$} & \textbf{\small{}$0.0195$} & \textbf{\small{}$0.6822$}\tabularnewline
\hline 
\textbf{\small{}$1000$} & \textbf{\small{}$0.0239$} & \textbf{\small{}$0.8084$} & \textbf{\small{}$0.0252$} & \textbf{\small{}$0.5129$}\tabularnewline
\hline 
\textbf{\small{}$5000$} & \textbf{\small{}$0.0296$} & \textbf{\small{}$0.8153$} & \textbf{\small{}$0.0362$} & \textbf{\small{}$0.3039$}\tabularnewline
\hline 
{\small{}$10,000$} & {\small{}$0.0320$} & {\small{}$0.8252$} & {\small{}$0.0379$} & {\small{}$0.2478$}\tabularnewline
\hline 
{\small{}$50,000$} & {\small{}$0.0402$} & {\small{}$0.8541$} & {\small{}$0.0435$} & {\small{}$0.1599$}\tabularnewline
\hline 
{\small{}$100,000$} & {\small{}$0.0482$} & {\small{}$0.8645$} & {\small{}$0.0494$} & {\small{}$0.1367$}\tabularnewline
\hline 
\end{tabular}\bigskip{}
\par\end{center}

\section{Conclusion}

\noindent In the present study, we have proposed an approach to reduce
the idiosyncratic overfitting of risk-analytic models by ensuring
that, for a fixed number of parameters, the probability distributions
used to model frequencies and severities are reasonably versatile.
After developing the concept of versatility based on underlying components
of functional simplicity and adaptability, we derived a mathematical
measure of this property (in the form of a normalized Bayesian Fisher
information matrix) in (19). This measure was refined to account for
parameterization differences by requiring, in (26), that the relevant
PMF or PDF be expressed in its least-compressible form.

The three simple (1-parameter) severity models of Table 2 revealed
that, for relatively flat distributions, the versatility measure increases
as the distribution is better able to concentrate probability around
its mode. Subsequently, we explored the implications of this measure
for a variety of 2-parameter severity and frequency distributions
in Tables 3 and 4, respectively. These examples showed that:
\begin{itemize}
\item The versatility measure of a 2-parameter distribution often, but not
always, falls between the measures of two 1-parameter special cases
(with each of the parameters fixed in turn), suggesting that the characteristics
contributed by the two parameters individually tend to be ``averaged''
when taken together.
\item Parameters with a direct effect on tail behavior \textendash{} for
example, $\tau$ for the $\textrm{Weibull}\left(\lambda,\tau\right)$
and $\textrm{Discrete}$\linebreak{}
$\textrm{Weibull}\left(\tfrac{m}{m+1},\tau\right)$ distributions
and $\alpha$ for the $\textrm{Pareto 2}\left(\alpha,\vartheta\right)$
and $\textrm{Waring}\left(\alpha,\vartheta\right)$ distributions
\textendash{} appear to have relatively larger impacts on versatility
than other parameters.
\item The versatility measures associated with certain discrete families
tend to be lower than those of their continuous analogues because
versatility is sensitive to the level of functional adaptability afforded
in the neighborhood of $0$.
\item The versatility measures of certain distributions (i.e., $\textrm{Pareto 2}\left(\alpha,\vartheta\right)$,
$\textrm{Lognormal}\left(\ln\left(\nu\right),\sigma=1\right)$,\linebreak{}
$\textrm{Negative Binomial}\left(r,\tfrac{m}{m+1}=\tfrac{1}{2}\right)$,
and $\textrm{Waring}\left(\alpha=1,\vartheta\right)$) are noticeably
lower than those of other distributions, raising serious concerns
about idiosyncratic behavior.
\end{itemize}
\qquad{}Finally, by comparing the $\textrm{Pareto 2}\left(\alpha,\vartheta\right)$
model with the more versatile $\textrm{Inverse Gamma}\left(r,\lambda\right)$
alternative, we showed how the proposed versatility measure may be
employed to assess the usefulness of probability distributions a priori,
effectively recognizing and mitigating problems of idiosyncratic overfitting.
Although our analysis does not propose minimum cutoff values for model-selection
purposes, we believe that, by observing large numbers of versatility
measures in applied work, it will be possible to establish reasonable
empirical guidelines for acceptable ranges of $\mathcal{V}\left(F_{0}\right)$.

Much further research remains to be done on this topic, particularly
involving the analysis of marginal contributions to versatility of
additional parameters for $k\geq2$. In this regard, it seems likely
that the most useful potential applications of the proposed versatility
measure lie in the investigation and comparison of probability distributions
within: (i) multiple-parameter continuous families, such as the Generalized
Beta, Burr, Pareto, and Pearson distributions used to model severities;
and (ii) discrete families modified to fit individual parameters to
specific values within the sample space (especially $x=0$ or $1$),
such as the Zero-Inflated Poisson and Negative Binomial models for
frequencies.

\end{document}